\documentclass[a4paper,11pt]{article}
\pdfoutput=1 

\usepackage{jheppub}

\usepackage[ddmmyyyy,hhmmss]{datetime}

\usepackage{amsmath}
\usepackage{amsfonts}
\usepackage[retainorgcmds]{IEEEtrantools}
\usepackage{tikz-feynman}
\usepackage{float}
\usepackage{verbatim}
\usepackage{slashed}

\hypersetup{pdfstartview={XYZ null null 0.75}}
\allowdisplaybreaks

\makeatletter \@addtoreset{equation}{section} \makeatother

\newcommand{\OO}{\mathcal{O}}
\newcommand{\LL}{\mathcal{L}}

\newcommand{\I}{\mathcal{I}}

\newcommand{\Y}{\mathbb{Y}}
\newcommand{\N}{\mathcal{N}}

\newcommand{\G}{\mathcal{G}}

\newcommand{\K}{\mathcal{K}}

\usepackage{amsmath,graphicx,amssymb,subfigure,bbm,comment}
\usepackage[all]{xy}
\usepackage{mathtools}
\usepackage{amsfonts}
\usepackage{comment}
\usepackage{mdframed}

\usepackage{feynmp}
\tikzfeynmanset{compat=1.1.0}

\usepackage{xcolor}
\usepackage{pifont}

\allowdisplaybreaks

\def\be{\begin{equation}}
\def\ee{\end{equation}}
\def\ba{\begin{eqnarray}}
\def\ea{\end{eqnarray}}

\def\a{\alpha}

\def\G{\Gamma}
\def\d{\delta}
\def\D{\Delta}

\def\s{\sigma}

\def\IR{\relax{\rm I\kern-.18em R}}

\def\IR{\relax{\rm I\kern-.18em R}}
\def\inv{^{\raise.15ex\hbox{${\scriptscriptstyle -}$}\kern-.05em 1}}

\title{Holography of a novel codimension-2 defect CFT}

\author[a]{George Georgiou}
\author[b,1]{, Georgios Linardopoulos, \note{Also at the Asia Pacific Center for Theoretical Physics (APCTP), Hogil Kim Memorial Building, \#501 POSTECH, 77 Cheongam-Ro Nam-gu, Pohang Gyeongsangbuk-do 37673, Korea.}}
\author[c]{and Dimitrios Zoakos}

\affiliation[a]{Agios Georgios Pagon, 49083, Corfu, Greece}
\affiliation[b]{Shanghai Institute for Mathematics and Interdisciplinary Sciences (SIMIS), \\ 657 Songhu Road, Yangpu District, Shanghai 200433, China.}
\affiliation[c]{Department of Physics, University of Patras, 26504 Patras, Greece.}

\emailAdd{georgios.georgiou2@gmail.com}
\emailAdd{george.linardopoulos@simis.cn}
\emailAdd{dzoakos@upatras.gr}

\abstract{We propose and study a new holographic duality between a non-supersymmetric defect conformal field theory (dCFT) and its gravity dual. On the gravity side, the defect is realised by a novel solution of a D5 probe brane embedded in the $AdS_5\times S^5$ geometry. The D5 brane wraps an $S^2\subset S^5$ and carries $k$ units of flux through the $S^2$. The symmetry of the induced on the brane metric is $AdS_3\times S^1\times S^2$. The brane ends on an $\mathbb{R}^{(1,1)}$ subspace of the 4-dimensional $AdS_5$ boundary resulting to a codimension-2 defect. We first prove that our brane configuration is stable by showing that the masses of all the fluctuations of the transverse to the brane coordinates respect the B-F bound.
On the field theory side, the 2-dimensional defect is described by a
classical solution whose precise form we determine. Subsequently, we calculate the one-point functions of the energy-momentum tensor and of the chiral primary operators (CPOs), first at strong and then at weak coupling. In an appropriate limit, we find compelling agreement between the weak and strong coupling results. Furthermore, we also extract one of the B-type Weyl anomaly coefficients.}

\begin{document}
\maketitle
\flushbottom


\section{Introduction}The dynamics of conformal field theories (CFTs) in the presence of defects describes a plethora of physical systems, ranging from boundaries and interfaces to objects such as Wilson loops, strings and branes. In general, the presence of a defect breaks, partially or even completely, the conformal symmetry of the ambient CFT rendering the calculation of observables much more involved. The situation becomes even more intricate if one is interested in the behaviour of the theory in the strong coupling regime. The calculations become rather tractable in the case where the ambient CFT is the maximally supersymmetric gauge theory in 4 dimensions, i.e ${\cal N}=4$ SYM. The reason is two-fold. On one hand, the ambient CFT is believed to be integrable in the planar limit \cite{Minahan:2002ve,Bena:2003wd} and on the other because of its holographic description in terms of type IIB string theory on $AdS_5\times S^5$ \cite{Maldacena:1997re}.

The defect CFT is characterised by whether the defect preserves some amount of supersymmetry \cite{Gaiotto:2008sa} and by whether the boundary conditions associated with the defect are integrable.
Another important characteristic of the defect is its codimension. The main focus of our work will be on codimension-2 defects in the context of gauge/gravity dualities. 
The most known and simplest defect of codimension-2 is the Gukov-Witten defect \cite{Gukov:2006jk}. Its holographic description is that of a probe D3 brane wrapping an $AdS_3\times S^1$ subspace of the $AdS_5\times S^5$ geometry. This defect preserves half of the supersymmetries, namely it respects a $\mathfrak{psu}(1,1|2)^2\rtimes \mathfrak{su}(2)_R $ subalgebra of the full superconformal algebra $\mathfrak{psu}(2,2|4)$. The corresponding field theory dual is described by classical solutions that have non-zero diagonal vevs for some of the scalar fields. These vevs exhibit a simple pole on the defect, a fact that is dictated by conformal invariance. At strong coupling, this defect is present in the classification of $1/2$-BPS integrable boundary conditions of the string sigma model presented in \cite{Dekel:2011ja} only for the special case  in which the probe brane has an inclination of $\pi/2$ with respect to the $AdS_5$
boundary.
The second $1/2$-BPS supersymmetric codimension-2 defect is the one preserving a $\mathfrak{su}(1,1|4)\times \mathfrak{su}(1,1)$ subalgebra of the $\mathfrak{psu}(2,2|4)$ superalgebra. It is, most likely, realised by a D7 brane wrapping an $AdS_3\times S^5$ subspace \cite{Harvey:2008zz}. This defect completes the list of $1/2$-BPS supersymmetric codimension-2 defects. For the case of non-supersymmetric defects which can be realised holographically, there is one that can be described by a flux stabilised 5-brane and which corresponds to a codimension-2 fuzzy $S^3$ in the classification of \cite{deLeeuw:2024qki}. 

In what follows, we present a case that apparently does not fall into any of the aforementioned classes. Our main interest will be on the holographic description of the duality and we will barely touch upon questions regarding whether the theories appearing on the two sides of the duality are integrable.\footnote{Recently the integrability properties of the Gukov-Witten defects were studied in \cite{Holguin:2025bfe,Chalabi:2025nbg}. The conclusion is that while ordinary Gukov-Witten defects are not integrable except for special sub-sectors, the rigid Gukov-Witten defects are integrable, at the leading order, in a corner of their parameter space.} This issue is left for future work.\footnote{Because there is nonzero flux present, we need to employ the generalized integrability framework introduced in \cite{LinardopoulosZarembo21, Linardopoulos22, Linardopoulos25a}.} The plan of the paper is as follows. In section \ref{D5-sol}, we introduce the embedding of our D5 brane solution in the ambient $AdS_5\times S^5$ geometry. The solution carries $k$-units of flux through an $S^2\subset S^5$. The geometry of the D5 brane is $AdS_3\times S^1\times S^2$. It should be noted that the intersection of $AdS_3\times S^1$ with the boundary of $AdS^5$ creates the surface defect whose codimension is 2. Let us, also, mention that the $AdS_3\times S^1$ has an arbitrary inclination angle with respect to the boundary of $AdS_5$, as it happens for the supersymmetric D3-D3 system \cite{Drukker:2008wr}. However, there is a major difference. For our solution to exist and to be stable the inclination angle can not exceed a critical value. In section \ref{stability}, we examine the stability of our configuration. In particular, we show that the masses of all the fluctuations of the transverse to the D5 brane coordinates respect the B-F bound, as long as the single parameter $\s$ characterising the solution is less than or equal to one.

In section \ref{Tmunu}, we perform the holographic calculation of the one-point function of the stress-energy tensor. The result takes the form dictated by the residual conformal symmetry. Furthermore, we extract the Weyl anomaly coefficient $d_2$ that is associated with the pullback of the bulk Weyl tensor to the defect. In section \ref{CPOs}, we focus on the strong coupling calculation of the one-point functions of the chiral primary operators (CPOs) of the ambient theory ${\cal N}=4$ SYM. Subsequently, in section \ref{dual} we conjecture which is the classical solution that describes the two dimensional interface generated by the D5 brane. Given this solution, we proceed to calculate the one-point function of the energy-momentum tensor at weak coupling. This calculation is presented in section \ref{T-weak}. Then, in section \ref{CPO-weak}, we perform the tree-level calculation for the case of the CPOs. Section \ref{agreement} is devoted to the comparison of the results between weak and strong coupling. We find that, there is a certain limit, namely that in which the square root of the 't Hooft coupling is much less than the flux through the 2-sphere $\frac{\sqrt{\lambda}}{k} \ll 1$, in which the weak and strong coupling results agree for both the energy-momentum tensor and the CPOs. This agreement provides strong evidence in favour of the
proposed duality. Finally, in section \ref{concl}, we present our conclusions and some possible future directions.



\section{The D5 probe brane}\label{D5-sol}

In this section, we introduce a novel D5-brane solution embedded in the $AdS_5\times S^5$ geometry. Our solution ends on a two-dimensional submanifold of the $AdS_5$ boundary and as such realises the holographic dual of a codimension-2 defect conformal field theory (dCFT). 
The D5-brane wraps an $S^2$ of the internal space $S^5$ and the symmetry of the induced, on the brane metric, is 
$AdS_3\times S^1\times S^2$.
The brane orientation of the codimension-2 D3-D5 probe-brane system 
(see \eqref{metric} for the metric and \eqref{embedding} for the embedding ansatz of the D5-brane) 
is given in table \ref{Table:C2D3D5system}, with the worldvolume coordinates of the D5-brane being $\zeta^\mu = (x_0,x_1,r,\psi, \beta, \gamma)$. 
\begin{table}[H]
\begin{center}\begin{tabular}{|c||c|c|c|c|c|c|c|c|c|c|}
\hline
& ${\color{red}x_0}$ & ${\color{red}x_1}$ & ${\color{red}r}$ & ${\color{red}\psi}$ & ${\color{red}z}$ & ${\color{blue}\tilde \psi}$ & ${\color{blue} {\tilde \beta} }$ & ${\color{blue} {\tilde{ \gamma}}}$ & ${\color{blue}\beta}$ & ${\color{blue}\gamma}$ \\ \hline
\text{D3} & $\bullet$ & $\bullet$ & $\bullet$ & $\bullet$ &&&&&& \\ \hline
\text{D5 probe} & $\bullet$ & $\bullet$ & $\bullet$ & $\bullet$ & & &&& $\bullet$ & $\bullet$ \\ \hline
\end{tabular}
\caption{The D3-D5 intersection.\label{Table:C2D3D5system}}\end{center}
\end{table}


\subsection{Embedding ansatz of the probe} 
\label{emb_ansatz}

In what follows, we will present our solution for the D5-brane which realises the gravity dual of a non-supersymmetric 
codimension-2 defect conformal field theory in the case of Lorentzian signature.
Our starting point will be the action
\begin{equation}
\label{D5-Lor}
S_{L}=- \frac{T_5}{g_s}\Bigg\{\int d^6 \zeta \sqrt{-{\rm det}\, \mathcal P [g+2 \pi \a' F]}-
2 \pi \a' \int \mathcal P [ F\wedge C_4]\Bigg\} 
\end{equation}
where $T_5$ is the tension of the D5-brane and $\mathcal P$ denotes the pullback of the different spacetime fields on the worldvolume of the brane. For the brane tension and the string coupling we have the following definitions 
\begin{equation} \label{def-tension-coupling}
T_5 = \frac{\lambda^{3/2}}{(2 \, \pi)^5} \quad \& \quad g_s = \frac{g_{YM}^2}{4\, \pi} \quad {\rm with} \quad \lambda = \alpha'^{-2} \, . 
\end{equation}
Notice, that we work in units in which the radius of the $AdS_5$ is taken to 1. 
In \eqref{D5-Lor} we have taken into account that the $B$-field vanishes for the $AdS_5\times S^5$ solution of the type IIB supergravity equations.

\begin{figure}[h!]
 \centering
  \includegraphics[width=0.9\textwidth]{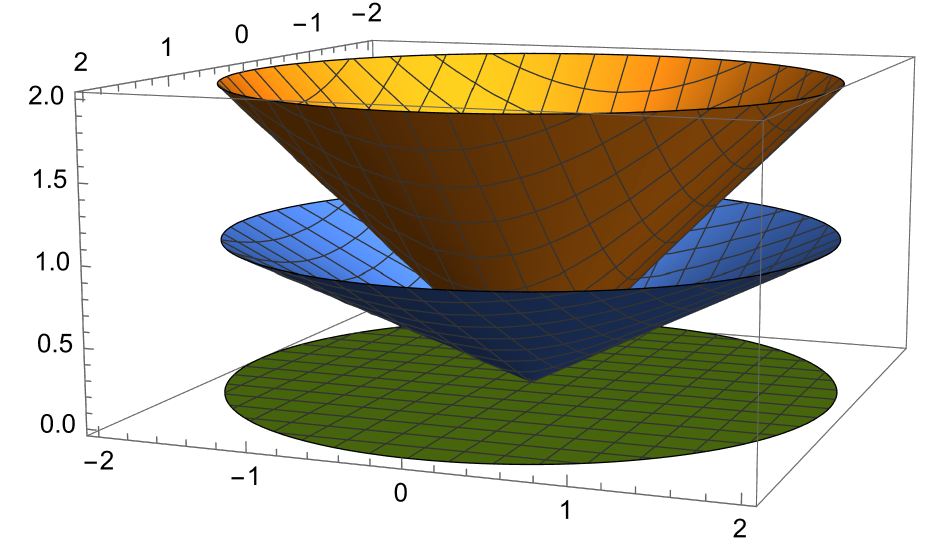}
 \caption{A picture depicting the D5 brane. The horizontal green plane parametrised by the coordinates $(x_2,x_3)=(r \cos{\psi},r \sin{\psi})$ represents the boundary of the $AdS_5$ space with the vertical axis being the holographic coordinate $z$. The blue surface represents the D5 brane while the orange is the critical surface $z=r$ after which the D5 brane becomes unstable (see eq. \eqref{stability2}). The point at which the brane touches the boundary is really a two dimensional surface since the coordinates $x_0$ ans $x_1$, along which the brane extends, have been suppressed.}
 \label{figg-1}
 \end{figure}

The complete background is given by
\begin{equation}\label{metric-Lor}
ds^2 = \frac{1}{z^2} \, \Big[-dx_0^2 + dx_1^2+ dr^2 + r^2 \, d\psi^2 + dz^2 \Big] +d\Omega_5^2
\end{equation}
with the metric of $S^5$ being
\begin{equation}\label{metric}
d\Omega_5^2 = 
d\tilde\psi^2 + 
\sin^2 \tilde\psi \, \left(d{\tilde \beta}^2 + \sin^2 {\tilde \beta} \, d{\tilde \gamma}^2 \right) +
\cos^2 \tilde\psi \left(d\beta^2 + \sin^2 \beta \, d\gamma^2\right) \, . 
\end{equation}
Additionally, the RR 4-form potential is given by
\begin{equation}\label{C4}
C_4 = - \frac{r}{z^4} \, dx_0 \wedge dx_1 \wedge dr \wedge d\psi \, . 
\end{equation}
As mentioned above, we choose the world-volume coordinates of a D5-brane to be $x_0,x_1,r,\psi,$ $ \beta, \gamma$ and consider the ansatz
\begin{equation}\label{embedding}
\tilde\psi = 0 \, , \quad 
{\tilde \beta} = \frac{\pi}{2} ,\quad 
{\tilde \gamma} =0 \quad \& \quad
z= \sigma \, r \, . 
\end{equation}
In figure \ref{figg-1} we depict the D5 brane with the coordinates $x_0, x_1, \beta$ and $\gamma$ suppressed. 
Let us make a brief comment on  the relation between
the parameter  $\s$ of our solution and the physical parameters $k$ and $\lambda$. In the limit of small $\s$  the flux $k$ becomes very large and the inclination of the brane with respect to the $AdS_5$ boundary goes to zero. On the other hand, when $\s=1$, which is the endpoint of the range of validity of our solution, the inclination angle becomes $\pi/4$ (see figure   \ref{figg-1}).

One can check that the ansatz above satisfies the equations of motion derived from the action \eqref{D5-Lor} 
once the following gauge field living on the world-volume of the D5-brane is turned on
\begin{equation}\label{A}
A = \frac{\kappa}{2 \, \pi \, \alpha'} \, \cos \beta \, d\gamma \quad {\rm with} \quad 
\kappa = \frac{4 + \sigma^2}{\sigma \sqrt{\big.8-\sigma^2}} 
\quad {\rm and} \quad 0<\sigma < 2\sqrt{2}
\end{equation} 
in units $\alpha' = \lambda^{-1/2}$. 
Let us mention that the equation of motion for the holographic coordinate determines the value of $\kappa$ in \eqref{A}, 
while the equation of motion for the world-volume field is used to determine its functional dependence. 
Given our ansatz, the equations of motion for the coordinates $\tilde \beta$, $\tilde \gamma$ 
and $\tilde\psi$ are satisfied automatically. 

At this point, let us comment on the codimension of the defect that our D5-brane introduces. Notice that the D5-brane intersects the $AdS_5$ boundary at $z=r=0$, implying that the defect is a two-dimensional plane $\mathbb R^{(1,1)}$ that extends along two of the four directions of the boundary, that is, along $x_0$ and $x_1$. As a result, we have a defect of codimension-2.

To quantise the flux we impose the following condition
\begin{equation}\label{kappa}
k=-\int_{S^2} \frac{F}{2 \, \pi} \quad \Rightarrow \quad 
k=\frac{1}{2 \pi}\frac{ \kappa}{2 \pi \alpha'}\int_0^\pi\sin \beta\,d\beta \int_0^{2\pi}d\gamma \quad \Rightarrow \quad k=\frac{\kappa}{\pi \a'}=\frac{\kappa \sqrt{\lambda}}{\pi}\, ,
\end{equation}
with $k \in {\mathbb N}^*$.
For the calculation of the one-point function of chiral primary operators, it is convenient to perform the Wick rotation 
($x^0 \rightarrow - i \, x^0$) in order to obtain the Euclidean version of the metric in \eqref{metric-Lor} and of the action in 
\eqref{D5-Lor}. The Euclidean action becomes the following
\begin{equation} \label{D5-Euclidean}
S_E= \frac{T_5}{g_s}\Bigg\{\int d^6 \zeta \sqrt{{\rm det}\, \mathcal P [g+2 \pi \a' F]}+i\, 
 2 \pi \a' \int \mathcal P [ F\wedge C_4]\Bigg\}\, ,
\end{equation}
where $C_4$ is still given by \eqref{C4}.

At this point, let us mention that we have performed the $\kappa$-symmetry analysis with the result being that our D5-brane solution breaks all supersymmetries, i.e.\ there is no Killing spinor of the $AdS^5\times S^5$ background that is preserved in the presence of our probe brane. This fact is in agreement with the dual field theory picture. The corresponding analysis in field theory is presented in appendix \ref{Appendix:Supersymmetry}.

Before closing this section let us write down the metric induced on the D5 brane world-volume. This reads
\be\label{induced-metric}
ds^2_{ind}=\frac{1}{r ^2 \s^2}\big(-dx_0^2+dx_1^2+(1+\s^2)dr^2\big)+\frac{1}{\s^2} d\psi^2+d\beta^2+ \sin^2\beta\, d\gamma^2.
\ee
After performing the change of variables $r=\frac{\hat r}{\sqrt{1+\s^2}}$ the metric above becomes
\be\label{induced-metric-1}
ds^2_{ind}=\frac{1+\s^2}{\s^2}\,\frac{-dx_0^2+dx_1^2+d\hat r^2}{\hat r^2}+\frac{1}{\s^2} d\psi^2+d\beta^2+ \sin^2\beta\, d\gamma^2\, .
\ee
In this last equation, the $AdS_3\times S^1\times S^2$ symmetry of the D5 brane becomes apparent.

\subsection{BF bound and stability of the solution}\label{stability}

For any calculation involving the defect probe brane to be meaningful, one must first prove the mechanical stability of the solution.
Consequently, in the present section we will focus on the stability of the defect D5-brane introduced in section \ref{emb_ansatz}.

We start by noticing that the symmetry of the metric induced on the D5-brane worldvolume is $AdS_3\times S^1\times S^2$, where $S^1$ corresponds to the isometry of the boundary $\psi$, while $S^2$ corresponds to the 2-sphere which our D5-brane wraps. As a result, the stability of the solution is ensured if the fluctuations of the coordinates transverse to the brane acquire masses which are above the Breitenlohner-Freedman (B-F) bound, namely $m^2\ge m^2_{BF}$ \cite{Breitenlohner:1982bm}. From the point of view of the D5-brane these fluctuations behave as scalars propagating on the hyperbolic space with geometry $AdS_3$. For a space-time with geometry $AdS_{d+1}$ the BF bound turns out to be $m^2_{BF}=-\frac{d^2}{4 \ell}$, where $\ell$ is the radius of $AdS_{d+1}$. 
In the units we are using, we have set $\ell=1$ and as a result our solution will be stable if the masses of the fluctuations satisfy $m^2\ge -1$. In what follows, we will see that this is the case, as long as the parameter of the solution $\sigma\le 1$. Thus, we see that the requirement of stability places a stronger constraint on $\s$ than the one put by the equations of motion.

We start by introducing fluctuations around the D5 solution both for the world-volume gauge field 
$A= \left\{\kappa \cos \beta + \d A(\zeta^{\mu})\right\} d\gamma $, as well as for transverse coordinates, i.e.\
\begin{equation}
z=\s \, r+ \d z(\zeta^{\mu})\, , \quad 
{\tilde \psi} = 0 + \d {\tilde \psi} (\zeta^{\mu})\, , \quad 
{\tilde \beta} = \frac{\pi}{2}+\d {\tilde \beta}(\zeta^{\mu}) \, , \quad 
{\tilde \gamma}= 0+\d {\tilde \gamma} (\zeta^{\mu}) \, . 
\end{equation}
The key idea is to plug these relations into the action \eqref{D5-Lor} and keep terms which are at most quadratic in the fluctuations.
In passing, let us mention that the terms linear in the fluctuations vanish due to the fact that we are expanding around a solution of the equations of motion. Then one uses this Lagrangian to
derive the equations governing the time evolution of the fluctuations from which one can determine the mass corresponding to each fluctuation and check if and under which conditions these masses satisfy the B-F bound.

Since we are primarily interested in the time evolution of the fluctuations, we further make the simplifying assumption that all fluctuations depend only on the time coordinate $x_0$. 
Then the equation of motion for $\d {\tilde \beta}(x_0)$ and $\d {\tilde \gamma}(x_0)$ are trivially satisfied while the equation of motion for $\d z(x_0)$ gives
\begin{equation} \label{dz}
\d z''(x_0)- \frac{1}{r^2} \,\d z(x_0)=0 
\end{equation}
where the prime denotes the derivative with respect to $x_0$ and we have used the relation between  $\kappa$  and the parameter $\s$ appearing in \eqref{A}. Inspecting \eqref{metric-Lor}, it is easy to go from time $x_0$ to proper time $s$ with $ds=\frac{dx_0}{z}=\frac{dx_0}{\s \,r}$. As a result \eqref{dz} becomes
\begin{equation}\label{dz1}
\d z''(s)- \sigma^2 \, \d z(s) =0
\end{equation}
where now the prime denotes the derivative with respect to $s$. We now make an ansatz for the fluctuation $\d z= c\, \sin{(m_z s)}$ which when inserted in \eqref{dz1} gives for the mass of the $z$ fluctuation the value
\begin{equation}\label{mz}
m_z^2 = - \, \s^2
\end{equation}
and the condition from the B-F bound
\begin{equation}\label{stability2}
m_z^2\, \ge \, - \, 1 \quad \Rightarrow \quad 
\s \le 1
\end{equation}
puts a constraint that it is stronger than $\s \le 2\sqrt{2}$.
Similarly, the equation of motion for $\d {\tilde \psi}$ 
when written in terms of $s$ reads
\begin{equation}
\d {\tilde \psi}''(s) + m_{\tilde \psi}^2 \,\d {\tilde \psi} =0 
\quad {\rm with} \quad 
m_{\tilde \psi}^2=-\frac{\s^2 (8 - \s^2)}{8 (1 + \s^2)} 
\end{equation}
and the corresponding mass is always greater than the BF bound, i.e.\ $m_{{\tilde \psi}}^2\ge -1$. Thus, the ${\tilde \psi}$ coordinate does not put any further constraint on $\s$.
Finally, for the fluctuation of the worldvolume gauge field $\d A$ one obtains the equation
\be
\d A''(s)=0 \Rightarrow m^2_A=0>-1\equiv m^2_{BF}.
\ee
In conclusion, we have seen that our D5-brane is stable as long as $\s\in (0,1]$.


\section{Holographic one-point function of the stress-energy tensor and anomaly coefficients}\label{Tmunu}

In this section, we present the holographic calculation of the one-point function of the energy momentum operator in the presence of the codimension two defect CFT. This important observable is zero both for the theory without any defect, as well as for any theory of a defect of codimension one. 

In the case of a codimension two defect CFT the residual conformal invariance constrains the space-time structure of the one point function to be fixed up to a function ${\mathbf h}={\mathbf h}(g_{YM},N)$
\be\label{1-pointCFT}
\langle T_{\mu\nu}\rangle={\mathbf h} \frac{\eta_{\mu\nu}}{r'^4}, \qquad \langle T_{ij}\rangle=\frac{{\mathbf h}}{r'^4}(4\, n_i\, n_j-3 \,\d_{ij}), \qquad \qquad \langle T_{\mu i}\rangle=0
\ee
In \eqref{1-pointCFT}, $r'$ denotes the radial distance of the position of the energy-momentum tensor from the planar surface defect. Furthermore, $x^\mu,\, \mu=0,1$ are the coordinates along the surface defect operator $\Sigma=R^2$ which is embedded in $R^4$, while $n_i=\frac{x^i}{r'},\, i=2,3$ is the unit vector normal to the surface defect.
Without any loss of generality, one can situate the energy-momentum operator $T_{\mu\nu}$ at the point of the boundary with coordinates $x^m=\{ x_0,x_1,x_2,x_3 \}=\{ 0,0,0,r' \}$.
With such a choice, the only non-zero components of the vacuum expectation of the energy momentum tensor are
\begin{equation} \label{nonzeroT}
\langle T_{00}\rangle=\langle T_{11}\rangle=\langle T_{33}\rangle= \frac{{\mathbf h}}{r'^4}, \qquad \langle T_{22}\rangle=-3 \frac{{\mathbf h}}{r'^4}
\end{equation}

From \eqref{1-pointCFT} it is straightforward to verify that the trace of the energy-momentum tensor is 0, namely $\langle T_{\,\,\, m}^m\rangle=0$. This happens because both the spacetime in which the defect is embedded, as well as the defect itself are flat. For the generic case of a 2-dimensional defect parametrised by $y^a,\,a=1,2$ with its embedding in a $D$-dimensional space described by $X^m(y),\, m=0,1,\cdots D-1$ the Weyl anomaly associated with the defect is given by~\cite{Graham:1999pm,Henningson:1999xi,Schwimmer:2008yh},
\be
\label{eq:defect-A}
\langle T^{m}_{~m} \rangle_{\textrm{def}} =-\frac{1}{24\pi}\left( b \, {\cal{R}}_{def} + d_1 \, Y^{\mu}_{ab}Y_{\mu}^{ab} - d_2 \, W_{ab}^{~~ab} \right),
\ee
where ${\cal{R}}_{def}$ is the defect's intrinsic scalar curvature, $W_{abcd}$ is the pullback of the bulk Weyl tensor to the defect. Furthermore, $Y^{\mu}_{ab}$ is the traceless part of the second fundamental form, namely $Y^{\mu}_{ab}\equiv \Pi^{\mu}_{~ab} - \frac{1}{2}\gamma_{ab} \gamma^{cd}\Pi^{\mu}_{~cd}$ with $\Pi^{\mu}_{~ab} = \hat{\nabla}_a \partial_b X^{\mu}$.\footnote{Here $\gamma_{ab}$ is the induced metric on the defect and $\hat{\nabla}_a$ the corresponding covariant derivative.} Finally, $b$, $d_1$, and $d_2$ are the defect central charges. Note that the codimension-2 Weyl anomaly generally includes two B-type parity odd terms which we omit. As discussed in \cite{Lewkowycz:2014jia,Jensen:2018rxu}, one can relate the anomaly coefficient $d_2$ to the value of the parameter ${\mathbf h}$ appearing in \eqref{1-pointCFT}. The precise relation is 
\be\label{hd2rel}
{\mathbf h} = -\frac{1}{2 \pi}\frac{1}{3\text{vol}(\mathbb{S}^{D-3})} \, \frac{D-3}{D-1} \, d_2,
\ee
where $D$ is the number of dimensions in which the 2-dimensional defect lives. In our case $D=4$.

The function $\mathbf h$ contains the particularities of the specific defect CFT and in this section we will calculate it at strong coupling using holography. 
To this end, let us review the main ingredients of the holographic calculation.
The IIB supergravity field that is dual to the stress tensor $T_{ij}$ is the fluctuations of the AdS metric:
\begin{IEEEeqnarray}{l}
g_{mn} = \hat{g}_{mn} + \delta g_{mn},
\end{IEEEeqnarray}
where $\hat{g}_{mn}$ is the (background) metric of AdS$_{d+1}$ which contains the probe Dp-brane. From now on, and for the rest of this section, the indices i, j = 0, 1, 2, 3 take values along the
directions parallel to the AdS boundary, while the Latin indices m, n = 0, 1, 2, 3, 4 along
all directions of $AdS_5$. One should be careful not to confuse the indices $i,j$ that appear from now on with the $i,j$ appearing in \eqref{1-pointCFT}. In \eqref{1-pointCFT} $i,j=2,3$.

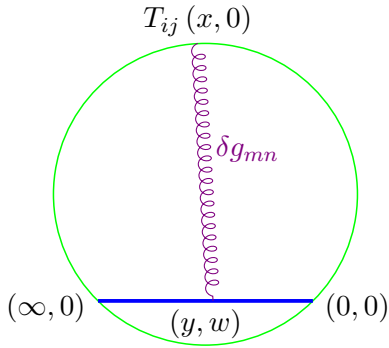
\begin{figure}[H]\begin{center}\begin{tikzpicture}
\draw[line width=0.2mm, green] (0,0) circle(2);\draw[line width=0.5mm, blue] (-1.414,-1.414) -- (1.414,-1.414);
\node (O3) at (-0.1,2.3) {$T_{ij}\left(x,0\right)$};
\node (O4) at (2,-1.5) {$\left(0,0\right)$};
\node (O5) at (-2.1,-1.5) {$\left(\infty,0\right)$};
\node (O6) at (0,-1.72) {$\left(y,w\right)$};
\begin{feynman}
\vertex (a) at (-.1,2);
\vertex (b) at (0.1,-1.414);
\diagram*{(a) -- [gluon, edge label=\(\delta g_{mn}\), violet] (b)};
\end{feynman}
\end{tikzpicture}\caption{Stress tensor one-point function in a defect CFT. The boundary of $AdS_5$ is represented by the green circle.}\label{Figure:StressTensorOnePointFunction}\end{center}\end{figure}

To compute the one-point function of the stress tensor in strongly coupled AdS/dCFT, we employ the recipe of \cite{Georgiou:2023yak}. The one-point function of the stress-energy tensor in the presence of a "heavy" object, here the Dp-brane, is given by \footnote{Whe the "heavy" object is the world-sheet of an operator with large dimension analogous computations can be found in \cite{Georgiou:2010an,Georgiou:2011qk}. }
\begin{IEEEeqnarray}{l}
\big\langle T_{ij}\left(x\right)\big\rangle_{\text{brane}} = \lim_{z\rightarrow 0}\big\langle\delta g_{ij}\left(x,z\right)\cdot \frac{1}{Z_{\text{brane}}}\int D\Y\,e^{-S_{\text{brane}}\left[\Y\right]}\big\rangle_{\text{bulk}}. \label{OnePointFunctionsDpBrane1}
\end{IEEEeqnarray}
The Dp-brane action $S_{\text{brane}}$ depends on the bulk supergravity modes $\delta g_{mn}$ via its dependence on the induced metric.\footnote{For more generic operators the fluctuations of the other supergravity background fields should also be taken into account.} By expanding the Dp-brane action around $\delta g_{mn} = 0$ we find,
\begin{IEEEeqnarray}{l}
S_{\text{brane}}\left[\Y\right] = S_{\text{brane}}\left[\Y\right]\Big|_{\delta g_{mn} = 0} + \left.\frac{\partial S_{\text{brane}}\left[\Y\right]}{\partial\delta g_{mn}}\right|_{\delta g_{mn} = 0}\cdot\delta g_{mn}\left(y,w\right) + \ldots \label{DpBraneAction}
\end{IEEEeqnarray}
\indent In the strong coupling regime ($\lambda\to\infty$), the path integral in \eqref{OnePointFunctionsDpBrane1} is dominated by a saddle point which corresponds to classical solutions $\Y_{\text{cl}}$ describing the embedding of the Dp-brane in the $AdS_5\times S^5$ spacetime. The first term in the expansion \eqref{DpBraneAction} then cancels the partition function which shows up in the denominator of \eqref{OnePointFunctionsDpBrane1} and the one-point function becomes:
\begin{IEEEeqnarray}{l}
\big\langle T_{ij}\left(x\right)\big\rangle_{\text{brane}} = -\lim_{z\rightarrow 0}\big\langle\delta g_{ij}\left(x,z\right)\cdot\left(\left.\frac{\partial S_{\text{brane}}\left[\Y_{\text{cl}}\right]}{\partial\delta g_{mn}}\right|_{\delta g_{mn} = 0}\cdot\delta g_{mn}\left(y,w\right)\right) \big\rangle_{\text{bulk}}, \label{OnePointFunctionsDpBrane2}
\end{IEEEeqnarray}
where the point $\left(x,0\right)$ lies on the boundary of AdS and denotes the position where the energy-momentum tensor sits, while $\left(y,w\right)$ is a point on the Dp-brane (see figure \ref{Figure:StressTensorOnePointFunction}). We have also used the fact that the boundary value of the one-point function vanishes, i.e.\ $\big\langle\delta g_{ij}\left(x\right)\big\rangle = 0$. 
At this point, let us stress once more that the indices $i,j=0,1,2,3$ take values along the directions parallel to the $AdS$ boundary, while the Latin indices $m,n=0,1,2,3,4$ along all directions of $AdS_5$.

The generic variation of a probe Dp-brane action with respect to the AdS metric $g_{mn}$ is given by
\begin{IEEEeqnarray}{l}
\frac{\partial S_{\text{brane}}\left[\Y\right]}{\partial\delta g_{mn}} = \frac{T_p}{2g_s}\int \left[d^{p+1}\zeta \sqrt{h} \, h^{ab} \partial_a \Y^{m} \partial_b \Y^{n}\right], \label{DBIvariation}
\end{IEEEeqnarray}
where $h_{ab}$ is the induced metric of the Dp-brane. Inserting the Dp-brane action variation \eqref{DBIvariation} into the one-point function formula \eqref{OnePointFunctionsDpBrane2} we are led to,
\begin{IEEEeqnarray}{ll}\label{1T}
&\big\langle T_{ij}\left(x\right)\big\rangle_{\text{brane}} = -\frac{T_p}{2g_s} \cdot \lim_{z\rightarrow 0}\int \left[d^{p+1}\zeta \sqrt{h} \, h^{ab} \partial_a \Y^{m} \partial_b \Y^{n} \right] \big\langle\delta g_{ij}\left(x,z\right) \cdot\delta g_{mn}\left(y,w\right)\big\rangle_{\text{bulk}} = 
\nonumber \\[6pt]
&=- \frac{T_p}{2 \, g_s} \, \frac{d+1}{d-1} \, \int \left[d^{p+1}\zeta \sqrt{h} h^{ab} \partial_a \Y^{m} \partial_b \Y^{n} \right] \frac{2}{w^{2}} \, \K_d(y,w;x) \, \I_{m k}\left(\tilde r\right)\I_{n l}\left(\tilde r\right) \, \I_{klij} \, .
\label{OnePointFunctionsDpBrane3}
\end{IEEEeqnarray}
In order to derive \eqref{1T} we have used the 
bulk-to-boundary propagator for the graviton in AdS$_{d+1}$ in the so-called de Donder gauge \cite{Liu:1998ty, Arutyunov:1999nw} which reads \footnote{For the factor of 2 in the second line of \eqref{1T}, see appendix \ref{Appendix:graviton-fluct}. This propagator was also used in the calculation of 3-point functions of 2 heavy operators and the energy-momentum tensor. The results obtained are in complete agreement to the conformal Ward identities \cite{Georgiou:2013ff}.}
\begin{IEEEeqnarray}{l}
\big\langle\delta g_{ij}\left(x\right) \cdot\delta g_{mn}\left(y,w\right)\big\rangle = 2\,\frac{d+1}{d-1} \cdot \frac{1}{w^2} \times \K_d(y,w;x) \, 
\I_{m k}\left(\tilde r\right) \, \I_{n l}\left(\tilde r\right) \, \I_{klij}
\label{GravitonPropagator}
\end{IEEEeqnarray}
where in our case $d=4$ and we have defined
\begin{IEEEeqnarray}{ll}
\tilde r \equiv\left\{y_i - x_i,w\right\}, \qquad \tilde r^2 \equiv \tilde r_m\tilde r_n \delta_{mn},
\end{IEEEeqnarray}
while the inversion tensors $\I_{mn}$, $\I_{klij}$ are given by
\begin{IEEEeqnarray}{l}
\I_{mn}(\tilde r) \equiv \delta_{mn} - \frac{2\tilde r_{m}\tilde r_{n}}{\tilde r^2} \quad {\rm and} \quad
\I_{klij} \equiv \frac{1}{2}\left(\delta_{ki}\delta_{lj} + \delta_{kj}\delta_{li}\right) - \frac{1}{d}\,\delta_{kl}\delta_{ij}. \label{InversionTensors}
\end{IEEEeqnarray}
Furthermore
\begin{equation} \label{PropagatorBulkToBoundary2}
\K_{4}\left(x,z;y\right) = \frac{c_{4} \cdot z^{4}}{\big(z^2 + \left(x - y\right)^2\big)^{4}} \quad {\rm with} \quad
c_{4} \equiv \frac{\Gamma\left(4\right)}{\pi^{2}\Gamma\left(2\right)} \, .
\end{equation}
Since the AdS$_{d+1}$ Poincaré coordinates are split as $x_{m} \equiv \left\{x_i,z\right\}$ and $y_{m} \equiv \left\{y_i,w\right\}$, the Latin indices $m,n = 0,\ldots,d$ are set to denote the full set of AdS coordinates, while the Latin indices $i,j,k,l$ only take the values $i,j,k,l = 0,\ldots, d-1$. 
%
At this point, we should mention that, from \eqref{1T} on, all manipulations are done as if all coordinates are Euclidean and the metric is $\d_{mn}$.

To proceed with the calculation of the one-point function \eqref{1T}, we need the expressions of the determinant of the induced metric on the D5-brane
and of the following quantity $h^{ab}\partial_a Y^\mu \partial_b Y^\nu \frac{\partial x'^{\kappa}}{\partial x^\mu} \frac{\partial x'^{\lambda}}{\partial x^\nu}$. 
Both are listed in the Appendix \ref{Appendix:details_computations},
in equations \eqref{det} and \eqref{D-brane-insertion}.
Notice, that a change of coordinates is needed, from $x^\mu=\{x_0, x_1, x_2, x_3, z\}$ to $x'^\kappa=\{x_0, x_1, r \cos{\psi}, r \sin{\psi}, z\}$, since the graviton propagator is written in Euclidean coordinates $x^\mu$ 
whereas the D5-brane solution in the polar coordinates $x'^\kappa$.

Now we should perform the 6-dimensional integral of \eqref{1T}. To do so, we first change variables from the Euclidean coordinates $(x_0,x_1)$ to polar ones $(\tilde r,\phi)$ related by 
$x_0= \tilde r \cos{\phi}$ and $x_1= \tilde r \sin{\phi}$. Subsequently, one can perform the integrals in the following order. First we perform the integral with respect to $\phi$, then the one with respect to $\tilde r$ followed by the $\psi$ and $r$ integrals. Finally, we perform the easy integrals with respect to the $S^2$ coordinates $\gamma$ and $\beta$.
In this way and by recalling that the operator is located at $\{ x_0,x_1,x_2,x_3 \}=\{ 0,0,0,r' \}$, one finally obtains the following expressions\footnote{To obtain the precise expressions for the components of the stress tensor, the definitions for the D5-brane tension and of the string coupling from \eqref{def-tension-coupling} have to be taken into account.}
\begin{equation}
\langle T_{22}\rangle = - \, 3 \, \langle T_{00}\rangle = 
- \, 3 \, \langle T_{11}\rangle= - \, 3 \, \langle T_{33}\rangle=- \, 3 
\Bigg[-\frac{ \lambda ^{3/2} \left(2 \sigma ^2+1\right)}{6 \pi ^3 \sigma ^3 \sqrt{8-\sigma ^2} g_{YM}^2}\Bigg] \, \frac{1}{r'^4 } \, . 
\end{equation}
This result has precisely the form of \eqref{nonzeroT} that is dictated by conformal invariance with 
\be\label{hT}
{\mathbf h}^{(strong)}=-\frac{ \lambda ^{3/2} \left(2 \sigma ^2+1\right)}{6 \pi ^3 \sigma ^3 \sqrt{8-\sigma ^2} g_{YM}^2}=-\frac{ \lambda ^{1/2} N \left(2 \sigma ^2+1\right)}{6 \pi ^3 \sigma ^3 \sqrt{8-\sigma ^2} }\, ,
\ee
where, as usual, $N$ is the number of colours.
As a bonus of our calculation from \eqref{hT} one can read the anomaly coefficient $d_2$ from \eqref{hd2rel}, which acquires the form
\be\label{d2}
d_2=\frac{6\, \lambda ^{3/2} \left(2 \sigma ^2+1\right)}{\pi \sigma ^3 \sqrt{8-\sigma ^2} g_{YM}^2}>0\, .
\ee
This result is in agreement with the fact that, if in the presence of the defect the averaged null energy condition holds, the coefficient $d_2$ should be nonnegative.
Finally, notice the scaling of the anomaly coefficient with respect to the 't Hooft coupling $\lambda$. This is essentially dictated by the dimensionality of the probe brane.


\section{Holographic one-point function of chiral primary operators}\label{CPOs}

In this section we present the holographic calculation of one of the most important data in defect CFTs, namely the calculation of the one-point function of chiral primary operators (CPOs).
Our calculation follows that of \cite{Nagasaki:2012re}.

The one-point function is non-zero only when the CPO respects the $SO(3)\times SO(3)$ symmetry of the defect brane. 
The corresponding spherical harmonic with the aforementioned symmetry can be found in \cite{Nagasaki:2012re,Kristjansen:2012tn} and is given by
\begin{align} \label{sph}
Y_\Delta(\tilde\psi)  
&= \frac{    (2+\Delta)! }{  2^{(\Delta+1)/2} \sqrt{(\Delta+1)(\Delta+2)}  }
\sum_{p=0}^{\Delta/2} \frac{(-1)^p \sin^{\Delta-2p}\tilde\psi
\cos^{2p}\tilde\psi }{ (2p+1)!(1+\Delta-2p)! } \, . 
\end{align}
The normalisation of the spherical harmonic and its value at $\tilde\psi=0$ are
\begin{equation} \label{def_Y_Delta}
\int_{S^5}  |Y_\Delta(\tilde\psi)|^2 =
\frac{1}{2^{\Delta-1}(\Delta+1)(\Delta+2)}\int_{S^5} 1
\quad \& \quad 
Y_\Delta(0)  
= \frac{(-1)^{\Delta/2}}{ 2^{(\Delta+1)/2}}
\sqrt{\frac{\Delta+2}{\Delta+1}} \, . 
\end{equation}
\begin{figure}[H]\begin{center}\begin{tikzpicture}
\draw[line width=0.2mm, green] (0,0) circle(2);\draw[line width=0.5mm, blue] (-1.414,-1.414) -- (1.414,-1.414);
\node (O3) at (-0.1,2.3) {${\cal O}_{\D}\left(X^\mu\right)$};
\node (O4) at (2,-1.5) {$\left(0,0\right)$};
\node (O5) at (-2.1,-1.5) {$\left(\infty,0\right)$};
\node (O6) at (0,-1.72) {$x^\mu(\zeta)$};
\begin{feynman}
\vertex (a) at (-.1,2);
\vertex (b) at (0.1,-1.414);
\diagram*{(a) -- [boson, edge label=\(\), violet] (b)};
\end{feynman}
\end{tikzpicture}\caption{One-point function of a CPO in a defect CFT. The CPO is located at the boundary point $X^\mu=(X_0,X_1,r',\psi',0)$ while $x^\mu(\zeta)$ denote the spacetime coordinates of an arbitrary point on the D5 brane which is parametrised by $\zeta=(x_0,x_1,r,\psi,\beta,\gamma)$ . Finally, the wavy line represents the fluctuations of the metric  and of the 4-form potential.}\label{Figure:OnePointFunction CPOs}\end{center}\end{figure}
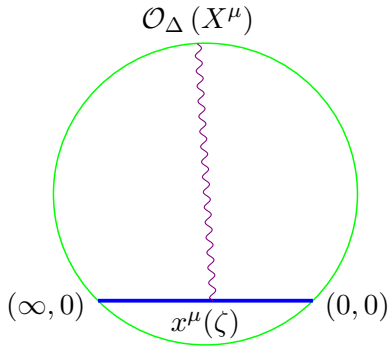
To calculate the one point function, we have to perturb the Euclidean version
of the action that is presented in section \ref{emb_ansatz}. 
The first order fluctuation of \eqref{D5-Euclidean} is
\begin{equation} \label{perturbed-DBI-WZ}
S^{(1)} = \frac{T_5}{g_s} \int d^6 \zeta \left({\cal L}^{(1)}_{DBI}  +i\, 
{\cal L}^{(1)}_{WZ}\right) 
\end{equation}
with 
\begin{equation} \label{Fluct_DBI+WZ}
{\cal L}^{(1)}_{DBI} = \frac{1}{2} \, \sqrt{\det H} \, \left(H^{-1}_{sym}\right)^{a b}
\partial_a X^M \, \partial_b X^N \, h_{MN}    \quad \& \quad 
\frac{{\cal L}^{(1)}_{WZ} }{2 \, \pi \alpha'}=F_{\beta \gamma} \, \frac{1}{4!}\epsilon^{abcd} (Pa)_{abcd}
\end{equation}
where $h_{\mu \nu}$ and $a_4$ are the fluctuations of the metric and of the RR 4-form in terms of the supergravity field $s$ which is dual to the CPO
\begin{align} \label{Fluct_components_metric_RR}
&h^\text{AdS}_{\mu\nu} 	=-\frac{2 \, \Delta \, (\Delta-1)}{\Delta+1}sg_{\mu\nu}+\frac{4}{\Delta+1}\nabla_\mu\nabla_\nu s
\nonumber \\[5pt]
&h^\text{S}_{\alpha\beta}	=2 \, \Delta \, s \, g_{\alpha\beta} 
\quad \& \quad
a^\text{AdS}_{\mu\nu\rho\sigma} = 4\, i \, 
\sqrt{g^\text{AdS}} \, \epsilon_{\mu\nu\rho\sigma\eta} \, \nabla^\eta s \, . 
\end{align}

The first order fluctuation of the DBI term from \eqref{Fluct_DBI+WZ}, 
for the specific D5-brane embedding ansatz \eqref{embedding}, becomes
\begin{eqnarray}
{\cal L}^{(1)}_{DBI}
&= & \frac{  \sin \beta}{8 \, r^3 \, \sigma^3 \, \sqrt{8 - \sigma^2}} 
 \Bigg[ 16 \, r^2 (1+\sigma^2)\left(h_{00} +h_{11}\right) + 16\, r^2 \, h_{22} + 16  (1+\sigma^2) h_{33}
\nonumber \\[5pt] 
&& + \, 32 \, r^2 \, \sigma \, h_{24} + 16 \, r^2 \, \sigma^2 \, h_{44} + (8-\sigma^2) 
\left(h_{88} + \frac{h_{99}}{\sin^2 \beta}\right) \Bigg]
 \\[5pt] 
 &= & \frac{- \, \Delta \, s \, \sin \beta}{2 \, r^3 \, \sigma^3 \, \sqrt{8 - \sigma^2}} 
 \Bigg[40+ \frac{16 \, \sigma^2}{K^2} \left[\left(x_0-X_0\right)^2 +\left(x_1-X_1\right)^2 + r^2 (1+\sigma^2) + r'^2\right]^2
\nonumber \\[5pt] 
&& +\,  \frac{32}{\sigma^2} + 17 \, \sigma^2  - 
\frac{32\,  (1+\sigma^2)}{K} \left[\left(x_0-X_0\right)^2 +\left(x_1-X_1\right)^2 +
 r^2 (1+\sigma^2) + r'^2\right] \Bigg]
\nonumber
\end{eqnarray}
where in the second equality of the equation above, we have substituted the values for the metric fluctuations from \eqref{metric_fluctuations}.
At this point let us mention that the CPO is located at the 
boundary point $(X_0,X_1,r',\psi',z=0)$. 
In figure \ref{Figure:OnePointFunction CPOs}, the positions of the CPO and 
of the  D5-brane are presented.
The next step is to substitute the value of $s$ from  \eqref{s_def} and integrate with respect to $x_0$ and $x_1$.
To do this we change variables as
\begin{equation} \label{R_Omega_change_variables}
x_0 - X_0 = R \, \cos \Omega \quad {\rm and } \quad x_1 - X_1 = R \, \sin \Omega
\quad {\rm with} \quad R \in [0,\infty)  \quad \& \quad \Omega \in [0,2\pi) \, . 
\end{equation}
Performing the integration along $R$ and along the angles $\beta$, $\gamma$ and $\Omega$, we arrive at the following expression.
\begin{equation} \label{DBIcontr_1}
\int {\cal L}^{(1)}_{DBI} \, d\beta\, d\gamma \, dx_0 \, dx_1 = 
\frac{2\, \pi^2  \, (r \, \sigma)^{\Delta-3}}{(1+\Delta) \, \sqrt{8-\sigma^2}}
\Bigg[ \frac{\Xi_1}{W^{\Delta-1}} +  \frac{\Xi_2\, \Upsilon}{W^{\Delta}}+  \frac{\Xi_3\, \Upsilon^2}{W^{\Delta+1}} \Bigg] c_{\Delta}\, Y_{\Delta}(0) 
\end{equation}
with
\begin{equation}
W = r^2 (1+\sigma^2) + r'^2 - 2 \, r \, r' \, \cos (\psi - \psi') \, , 
\quad 
\Upsilon = r^2 (1+\sigma^2) + r'^2
\end{equation}
and with the values for the three parameters $\Xi_1$, $\Xi_2$ and $\Xi_3$ being 
\begin{eqnarray}
\Xi_1 &=& \frac{8 \left(5 \Delta ^2+\Delta -4\right) \sigma ^2+\Delta  (17 \Delta -15) 
\sigma ^4+32 \Delta  (\Delta +1)}{(1-\Delta ) \sigma ^2}
\\[5pt] 
\Xi_2 &=& 32 \left(\Delta  \sigma ^2+\Delta +1\right) \, , 
\quad 
\Xi_3 =- \, 16 \, \Delta \,  \sigma ^2 \, . 
\nonumber
\end{eqnarray}
To proceed we perform another change of variables from $r$ to $\zeta$, namely
\begin{equation}
\frac{2 \, r \, r'}{\Upsilon} = \sqrt{1- \frac{1}{\zeta^2}} 
\quad {\rm with} \quad \zeta \in \left[1,\sqrt{1+\frac{1}{\sigma^2}}\right] \, . 
\end{equation}
Now the DBI contribution
\begin{equation}
{\cal L}^{Contribution}_{DBI} = \int {\cal L}^{(1)}_{DBI} \, d\beta\, d\gamma \, dx_0 \, dx_1 \, dr\, d\psi 
\end{equation}
can be written as the following sequence of integrals
\begin{eqnarray}\label{DBI-contr}
{\cal L}^{Contribution}_{DBI} & =  & +  \, \frac{1}{r'^{\Delta}}
\frac{2^{2-\Delta}\, \pi^2  \, \sigma^{\Delta-3}}{(1+\Delta) \, \sqrt{8-\sigma^2}} 
\int_1^{\sqrt{1+\frac{1}{\sigma^2}}} d\zeta \, 
\frac{\left(\zeta^2-1\right)^{\frac{\Delta}{2}-2}}{\sqrt{\big.1+(1-\zeta^2)\sigma^2}}
c_{\Delta}\, Y_{\Delta}(0) 
\nonumber \\
&& \times  \Big[\zeta-  \sqrt{\big.1+(1-\zeta^2)\sigma^2}\Big] \Big[ \Xi_1 \, {\cal I}_{\Delta-1} 
+ \Xi_2 \, \zeta  \, {\cal I}_{\Delta} + \Xi_3 \, \zeta^2 \, {\cal I}_{\Delta+1} \Big]
\end{eqnarray}
where ${\cal I}_{\delta}$ denotes the following integral
\begin{equation}
{\cal I}_{\delta} =  \int_0^{2 \, \pi}
\frac{d \psi}{\Big[\zeta - \sqrt{\big. \zeta^2-1} \, 
\cos (\psi - \psi') \Big]^{\delta}} \, .  
\end{equation}
This class of integrals can be calculated 
by using the following integral representation of the associated Legendre polynomials
\begin{equation}
\int_0^{2\pi}  d\psi \frac{ \cos{k (\psi-\psi')}} {(\zeta-\sqrt{\zeta^2-1} \cos{ (\psi-\psi'))^\d}}=2 \pi \frac{(\d-1-|k|)!}{(\d-1)!} P_{\d-1}^{|k|}(\zeta)\, .
\end{equation}

Notice also that in \eqref{DBI-contr} only the first term in the first bracket of the second line of \eqref{DBI-contr} contributes, that is only the term with a branch cut in $\sqrt{\big.1+(1-\zeta^2)\sigma^2}$ contributes in the integral computation. This is so because when we make the change the variable from $r$ to $\zeta$ the resulting integral should be integrated from 1 to $\sqrt{1+\frac{1}{\sigma^2}}$ and back. But the integrand coming from the second term in the first bracket of the second line of \eqref{DBI-contr} is even in the $\zeta$ variable giving, thus, a zero result.

The explicit form of the associated Legendre polynomials is given by
\begin{equation}
P_{\d}^{|k|}(x)=\frac{(x^2 - 1)^{(|k|/2)}}{2^\d \d!} \, 
\frac{d^{\d+|k|} (x^2 - 1)^\d}{dx^{\d+|k|}} \, . 
\end{equation}
Finally, we have one last change of variables from $\zeta$ to $\chi$, that is,
\begin{equation}
\zeta^2 = 1+ \frac{1}{\sigma^2} \sin^2 \chi \quad {\rm with } \quad 0 \le \chi \le\frac{\pi}{2} \, . 
\end{equation}
The DBI contribution to the one-point function of the  single CPO with conformal dimension $\D$ becomes\footnote{There is an extra factor of 2 
in \eqref{Full_DBI}, since we integrate $\chi$ between $0$ and $\pi/2$ and not between $0$ and $\pi$.}
\begin{equation} \label{Full_DBI}
{\cal L}^{Contribution}_{DBI} = +  \, \frac{1}{r'^{\Delta}}
\frac{2^{4-\Delta}\, \pi^3}{(1+\Delta) \,\sigma\,  \sqrt{8-\sigma^2}} 
\Bigg[ \Xi_1 \, \Psi_1 +\Xi_2\, \Psi_2+ \Xi_3  \, \Psi_3 \Bigg]
c_{\Delta}\, Y_{\Delta}(0) 
\end{equation}
where with $\Psi_1$, $\Psi_2$ and $\Psi_3$ we denote the following integrals
\begin{equation} \label{Psi_1}
\Psi_1 = \int_0^{\frac{\pi}{2}} d \chi \, 
\sin^{\Delta -3} \chi \, 
P_{\Delta-2}^{(0)} \left(\sqrt{\big. 1+ \frac{1}{\sigma^2} \sin^2 \chi}\right)
\end{equation}
\begin{equation} \label{Psi_2}
\Psi_2 = \int_0^{\frac{\pi}{2}} d \chi \, 
\sqrt{\big.1+ \frac{1}{\sigma^2}\, \sin^2 \chi }\, 
\sin^{\Delta -3} \chi \, 
P_{\Delta-1}^{(0)} \left(\sqrt{\big. 1+ \frac{1}{\sigma^2} \sin^2 \chi}\right)
\end{equation}
\begin{equation} \label{Psi_3}
\Psi_3 = \int_0^{\frac{\pi}{2}} d \chi \, 
\Bigg[1+ \frac{1}{\sigma^2}\, \sin^2 \chi \Bigg]\, 
\sin^{\Delta -3} \chi \, 
P_{\Delta}^{(0)} \left(\sqrt{\big. 1+ \frac{1}{\sigma^2} \sin^2 \chi}\right) \, .
\end{equation}
Notice that $\Psi_1$, $\Psi_2$ and $\Psi_3$ cannot be calculated for generic values of the conformal dimension; however, for specific values of $\Delta$ the integrals can be performed.

The first order fluctuation of the WZ term proceeds similarly. Substituting the
specific D5-brane embedding ansatz of \eqref{embedding} into the
WZ term of \eqref{Fluct_DBI+WZ} and using the expression for the fluctuation of the RR four form from \eqref{Fluct_components_metric_RR}, we arrive at the following 
\begin{eqnarray}
{\cal L}^{(1)}_{WZ} &=& 4\, i \, \kappa \, \sin \beta \, \sqrt{g^{AdS}} \Big[ g^{zz} \partial_z - \sigma \, g^{rr} \partial_r\Big] s
\\[5pt] 
 &= &i\, \frac{4\, \Delta \, s \, \sin \beta \, (4+ \sigma^2)}{r^3 \, \sigma^5 \, \sqrt{8 - \sigma^2}} 
 \Bigg[1+ \sigma^2 -
\frac{\sigma^2}{K} \left[\left(x_0-X_0\right)^2 +\left(x_1-X_1\right)^2 +
 r^2 (1+\sigma^2) + r'^2\right] \Bigg] \, . 
\nonumber
\end{eqnarray}
Integrating along $\beta$, $\gamma$, $x_0$ and $x_1$ after the change of variables explained in \eqref{R_Omega_change_variables}, we arrive at the following expression
\begin{equation}
\int {\cal L}^{(1)}_{WZ} \, d\beta\, d\gamma \, dx_0 \, dx_1 = 
i\,\frac{16\, \pi^2  \, r^{\Delta-3} \, \sigma^{\Delta-5} \,  \left(4+\sigma^2\right)}{(\Delta-1) \, \sqrt{8-\sigma^2}}
\Bigg[ \frac{\Xi_4\, \Upsilon}{W^{\Delta}}+  \frac{\Xi_5}{W^{\Delta-1}} \Bigg]
c_{\Delta}\, Y_{\Delta}(0) 
\end{equation}
with
\begin{equation}
\Xi_4 = \left(1- \Delta \right) \sigma^2 \, , 
\quad 
\Xi_5 = \Delta - \Xi_4 \, . 
\end{equation}
Changing coordinates from $r$ to $\zeta$ we arrive to the following expression for the complete WZ contribution
\begin{equation}
{\cal L}^{Contribution}_{WZ} = i\,\int {\cal L}^{(1)}_{WZ} \, d\beta\, d\gamma \, dx_0 \, dx_1 \, dr\, d\psi 
\end{equation}
with
\begin{eqnarray}
{\cal L}^{Contribution}_{WZ} & =  & \,  \frac{1}{r'^{\Delta}}
\frac{2^{5-\Delta}\, \pi^2  \, \sigma^{\Delta-5} \, (4+\sigma^2)}{(\Delta-1) \, \sqrt{8-\sigma^2}} 
\int_1^{\sqrt{1+\frac{1}{\sigma^2}}} d\zeta \, 
\frac{\left(\zeta^2-1\right)^{\frac{\Delta}{2}-2}}{\sqrt{\big.1+(1-\zeta^2)\sigma^2}}
c_{\Delta}\, Y_{\Delta}(0) 
\nonumber \\
&& \times  \Big[\zeta-  \sqrt{\big.1+(1-\zeta^2)\sigma^2}\Big] \Big[ 
\Xi_4 \, \zeta  \, {\cal I}_{\Delta} + \Xi_5 \,\, {\cal I}_{\Delta-1} \Big] \, .
\end{eqnarray}
Similar to the DBI case, there is an extra factor of 2 since we integrate $\chi$ between $0$ and $\pi/2$. Performing the integrals along $\zeta$, considering only the terms with 
a branch cut at $\sqrt{\big.1+(1-\zeta^2)\sigma^2}$, we arrive to the following expression for the WZ contribution
\begin{equation} \label{Full_WZ}
{\cal L}^{Contribution}_{WZ} = \,   \frac{1}{r'^{\Delta}}
\frac{2^{7-\Delta}\, \pi^3  \, (4+\sigma^2)}{(\Delta-1) \, \sigma^3 \, \sqrt{8-\sigma^2}} 
\Bigg[ \Xi_4 \, \Psi_2 +\Xi_5\, \Psi_1 \Bigg] c_{\Delta}\, Y_{\Delta}(0) \, . 
\end{equation}
Notice that for the calculation of the WZ contribution, no new $\chi$ integrals have to be computed. We just need to recycle the $\Psi$ integrals from the DBI computation.

Substituting into \eqref{perturbed-DBI-WZ} the contributions of DBI \eqref{Full_DBI} and WZ \eqref{Full_WZ}, together with the definitions for the constants $c_{\Delta}$ \eqref{s_def} and $Y_{\Delta}(0)$ \eqref{def_Y_Delta}, we arrive at the following expression for the one-point function of the chiral primary operator
\begin{eqnarray}\label{O-general}
 \langle {\cal O}_\D(r')\rangle^{(strong)} &=&  \,    
 \frac{i^{\Delta } 2^{-\Delta -\frac{3}{2}}
   }{\pi \, \sqrt{\Delta }}\, \sqrt{\frac{\Delta +2}{\Delta +1}}\,
   \frac{\sigma }{\sqrt{8-\sigma ^2}} \,   \frac{\sqrt{\lambda }}{\Delta -1}  \, \frac{1}{r'^\D} \, 
\\
   &&\times  
   \Bigg[\left(9 \Delta ^2-15 \Delta +8\right)\,\Psi_1 -8 (\Delta -1) (3 \Delta -1) \Psi_2 +
   16 (\Delta -1) \Delta  \Psi_3\Bigg] \, .
   \nonumber
\end{eqnarray}
To derive \eqref{O-general} we have used the relation $\langle{\cal O}_\D(x)\rangle=-\,({\cal L}^{Contribution}_{DBI}+{\cal L}^{Contribution}_{WZ})$. The overall minus which originates from the usual relation $\langle{\cal O}_\D(x)\rangle=-\frac{\d S^{(1)}_E[s_0(x)]}{\d s_0(x)}$. The latter follows from the AdS/CFT prescription for the relation of the partition function of string theory to the generating functional in field theory, namely $\langle e^{ \int d^4x\, s_0(x){\cal O}_\D(x)} \rangle=e^{-S_E[s_0(x)]}$ \cite{Gubser:1998bc}.

Since the integrals $\Psi_1$, $\Psi_2$ and $\Psi_3$  cannot be calculated for a generic value of $\Delta$, we present the expression of the one-point function for specific values of the conformal dimension. For $\Delta=4$, $\Delta=6$ and $\Delta=8$ the one-point becomes\footnote{Notice that for the special case $\D=2$ the integrals in \eqref{O-general} diverge. This is similar to what happens in the D3-D7 system \cite{Kristjansen:2012tn}.}
\begin{eqnarray} \label{one-point-function-Delta-4-6-8}
  \langle {\cal O}_4(r')\rangle^{(strong)} & =  &  \frac{\sqrt{\lambda }}{16 \,  \sqrt{15} \, \pi } \, \frac{1}{r'^4}\,\frac{1+\sigma ^2}{\sigma ^5  \, \sqrt{8-\sigma ^2}}
   \, \left(5 \sigma ^4+56 \sigma
   ^2+96\right) \nonumber \\[5pt]
    \langle {\cal O}_6(r')\rangle^{(strong)} & = &  
    - \frac{\sqrt{\lambda }}{80 \, \sqrt{42} \, \pi }\, \frac{1}{r'^6} \, 
    \frac{1+\sigma ^2}{ 
   \sigma ^7 \, \sqrt{8-\sigma ^2}} \, 
   \left(7 \sigma ^6+165 \sigma ^4+672 \sigma ^2+640\right)
   \\[5pt]
    \langle {\cal O}_8(r')\rangle^{(strong)} & = & 
   \frac{\sqrt{\lambda }}{336 \, \sqrt{10} \, \pi }\, \frac{1}{r'^8}\frac{1+\sigma ^2}{\sigma ^9 \sqrt{8-\sigma ^2}}\, 
     \left(3 \sigma ^8+112 \sigma ^6+790 \sigma ^4+1720 \sigma ^2+1120 \right) \, . 
    \nonumber
\end{eqnarray}
At this point, let us mention that it is straightforward to evaluate the result for CPOs of arbitrary large conformal dimension $\D$.


\section{Conformal Field Theory dual and one-point functions at weak coupling}\label{dual}

In this section, we will discuss the details of the dual to our D5-brane defect CFT. Subsequently, we will present the weak coupling calculations of the one-point functions of the energy-momentum tensor and of chiral primary operators whose strongly coupled counterparts can be found in sections \ref{Tmunu} and \ref{CPOs}, respectively. 

We start by writing down the ${\cal N}=4$ SYM Lagrangian density in the mostly plus Lorentzian signature which reads
\begin{IEEEeqnarray}{ll}
\LL_{\N = 4} = \frac{2}{g_{\text{\scalebox{.8}{YM}}}^2} \text{tr}\bigg\{-\frac{1}{4} F_{\mu\nu} F^{\mu\nu} - \frac{1}{2} \left(D_{\mu}\varphi_i\right)^2 + \frac{i}{2}\,\bar{\psi}\slashed{D}\,\psi +\frac{1}{2}\,\bar{\psi}\,\Gamma^{i+3}\, \left[\varphi_i, \psi \right]+ \frac{1}{4}\left[\varphi_i,\varphi_j\right]^2\bigg\} \nonumber \\
\label{LagrangianSYM}
\end{IEEEeqnarray}
where, as usual, $\bar{\psi}_{\alpha} \equiv \psi_{\alpha}^{\dagger} \Gamma^0$, $\slashed{D} \equiv \Gamma^{\mu}D_{\mu}$ and the gauge field strength is 
\begin{IEEEeqnarray}{l}
F_{\mu\nu} \equiv \partial_{\mu}A_{\nu} - \partial_{\nu}A_{\mu} - i \left[A_{\mu},A_{\nu}\right], \quad D_{\mu}f \equiv \partial_{\mu}f - i \left[A_{\mu},f\right] \, .  \label{CovariantDerivatives}
\end{IEEEeqnarray}
Here $\Gamma^{M}$ and $\psi$ are the 10-dimensional gamma matrices and fermion fields, respectively. The matrices $\Gamma^M$ obey the 10-dimensional Clifford algebra, 
with $M=(\mu,i+3),\, \mu=0,1,2,3, \,i= 1,2,3,4,5,6$. The scalars $\varphi_i$ denote the six real scalar fields of $\mathcal{N}=4$ SYM. All fields transform in the adjoint representation of $\mathfrak{su}(N)$.

 The equations of motion for the bosonic fields derived from \eqref{LagrangianSYM} are the following:
\begin{IEEEeqnarray}{c}\label{eoms}
D^{\mu}F_{\mu\nu} = i\left[D_{\nu}\varphi_i,\varphi_i\right], \qquad D^{\mu}D_{\mu}\varphi_i = \left[\varphi_j,\left[\varphi_j,\varphi_i\right]\right]. 
\end{IEEEeqnarray}
For the case of a non-supersymmetric codimension-2 defect with zero gauge field $A_\mu=0$ the equations \eqref{eoms} reduce to
\begin{IEEEeqnarray}{c}\label{eoms1}
\left[\partial_{\nu}\varphi_i,\varphi_i\right]=0, \qquad \partial^{\mu}\partial_{\mu}\varphi_i = \left[\varphi_j,\left[\varphi_j,\varphi_i\right]\right].
\end{IEEEeqnarray}
Due to the symmetry of our D5-brane solution we can assume that the scalar fields
depend only on the radial distance from the defect $r'=\sqrt{x'^2_2+x'^2_3}$, namely $\varphi_i=\varphi_i(r')$. In such case the second equation in \eqref{eoms1} further simplifies to 
\be\label{radial}
\frac{d^2\varphi_i}{dr'^2}+\frac{1}{r'}\frac{d\varphi_i}{dr'}= \left[\varphi_j,\left[\varphi_j,\varphi_i\right]\right].
\ee
The last equation along with the first equation in \eqref{eoms1} admit the following solution which is the field theory dual of our probe D5 brane
\begin{IEEEeqnarray}{l}\label{sol}
\varphi_{i+3} = \varphi_{i+3}^{\text{cl}}\left(r'\right) = \frac{1}{r'} \cdot \left[\begin{array}{cc} \left(t_i\right)_{k\times k} & 0_{k\times \left(N - k\right)} \\ 0_{\left(N - k\right)\times k} & 0_{\left(N - k\right)\times \left(N - k\right)} \end{array}\right]_{N\times N} \\
\varphi_{i} = 0, \qquad i=1,2,3\, , \nonumber \label{FuzzyFunnelD3D5}
\end{IEEEeqnarray}
where the matrices $t_i$ realise a $k$-dimensional irreducible representation of $\mathfrak{su}\left(2\right)$ satisfying
\begin{IEEEeqnarray}{c}\label{tmatr}
\left[t_i, t_j\right] = \frac{i}{ \sqrt{2}}\epsilon_{ijl}t_l, \qquad i,j,l = 1,2,3.
\end{IEEEeqnarray}
For \eqref{sol} to be a solution of \eqref{radial} the matrices $t_i$ should satisfy 
\be\label{comm}
t_i=\sum_{j=1}^3\left[t_j,\left[t_j,t_i\right]\right]
\ee
which is fully consistent with \eqref{tmatr}.
In the dual gravity description, $k$ corresponds to the integer flux through the $S^2\subset S^5$ (see \eqref{kappa}). Let us stress that in our theory and in contradistinction to other defect CFT setups $k$ cannot be zero, the mere existence of our solution requires $k\neq 0$.
At this point, let us make some further comments on the symmetries of
the field theory solution of the current section. First notice that, on the field theory side, the conformal symmetry of the duality is manifest in the form of the solution \eqref{sol} that scales as $\frac{1}{r'}$. Furthermore, as discussed in section \ref{emb_ansatz}, the symmetry of the brane is $AdS_3\times S^1$. The existence of the isometry $S^1$, which is related to the angle $\psi$, implies that the expectation value of the scalar fields should not depend on the angle $\psi$. In contradistinction, in the case of the co-dimension 2 supersymmetric D3-D3 system the analogous solution depends explicitly on the angle $\psi$ since in this case the isometry is defined by the relation $\phi+\psi=\phi_0$ which relates the internal angle $\phi$ to the boundary angle $\psi$ \cite{Drukker:2008wr}. So the manifestation of the isometry $S^1$ is that it does not appear in the field theory ansatz. 
Finally, the choice of the above solution is dictated by the fact that in the dual gravity theory the D5-brane sits at $\tilde \psi=0$ which implies that only the three scalars related to one of the $S^2$s of \eqref{metric} should have non-zero vacuum expectation values (vevs), namely $\varphi_{i+3}, \,i =1,2,3$.


\subsection{One-point function of the energy-momentum tensor}\label{T-weak}
We now turn to the weak coupling calculation of the energy-momentum tensor. To this end we need the improved energy-momentum tensor derived from \eqref{LagrangianSYM}. 
The relevant for our purpose part is the part containing only scalars and this is given by \cite{CallanColemanJackiw70,deLeeuw:2023wjq, Linardopoulos25b}
\begin{IEEEeqnarray}{ll}
T_{\mu\nu}^{ \text{(scalars)}} = \frac{2}{g_{\text{\scalebox{.8}{YM}}}^2} \cdot \text{tr}\bigg\{&\frac{2}{3}\left(\partial_{\mu}\varphi_i\right)\Bigl(\partial_{\nu}\varphi_i\Bigr) - \frac{1}{3}\,\varphi_i\left(\partial_{\mu}\partial_{\nu}\varphi_i\right) - \nonumber \\
& - \frac{1}{6} \, \eta_{\mu\nu}\left[\left(\partial_{\varrho}\varphi_i\right)^2 + \frac{1}{2}\left[\varphi_i,\varphi_j\right]^2\right]\bigg\}, \qquad \label{StressTensorScalar}
\end{IEEEeqnarray}
where in this equation $i=1,2,\dots,6$.
One should now plug in \eqref{StressTensorScalar} the classical solution \eqref{sol} in order to find the one-point function of the energy-momentum tensor.
To this end, the following relations are being used
\be\label{rel1}
 \text{tr}(\partial_\mu \varphi_i\partial_\nu \varphi_i)= 
 \Bigg[\d_{\mu 2} \d_{\nu 2}\frac{x'^2_2}{r'^6}+\d_{\mu 3} \d_{\nu 3}\frac{x'^3_2}{r'^6}+(\d_{\mu 2} \d_{\nu 3}+\d_{\mu 3} \d_{\nu 2})\frac{x'_2x'_3}{r'^6}\Bigg]\cdot 
 \text{tr}\Big[\left(t_i\right)_{k\times k}\left(t_i\right)_{k\times k}\Big]
\ee
\begin{IEEEeqnarray}{ll} \label{rel2}
 \text{tr}( \varphi_i\partial_\mu\partial_\nu \varphi_i)&=
 \Bigg[-(\d_{\mu 2} \d_{\nu 2}+\d_{\mu 3} \d_{\nu 3})\frac{1}{r'^4} + \frac{3}{r'^6}\Big\{ \d_{\mu 2} \d_{\nu 2}x'^2_2+ \d_{\mu 3} \d_{\nu 3}x'^3_2+\nonumber\\
& (\d_{\mu 2} \d_{\nu 3}+\d_{\mu 3} \d_{\nu 2})x'_2 x'_3\Big\}\Bigg]\cdot 
 \text{tr}\Big[\left(t_i\right)_{k\times k}\left(t_i\right)_{k\times k}\Big]
\end{IEEEeqnarray}
and
\begin{equation}\label{rel3}
 \text{tr}(\left[\varphi_i,\varphi_j\right]^2)=-\frac{1}{r'^4} \cdot
 \text{tr}\Big[\left(t_i\right)_{k\times k}\left(t_i\right)_{k\times k}\Big],
\end{equation}
where for the last equation we have used twice \eqref{tmatr}.
To avoid confusion, let us mention that in the left hand side of \eqref{rel1}, \eqref{rel2} and \eqref{rel3}, $i$ and $j$ run from 1 to 6 while in the right hand side of the same equations $i$ runs from 1 to 3
in accordance to \eqref{tmatr}.

Finally, using the following useful relation of the matrices $t_i$ 
\be\label{normm}
\sum_{i=1}^3 (\varphi_{i+3}^{\text{cl}})^2=\frac{1}{8 r'^2}(k^2-1) {\mathbf 1}_{k\times k} \oplus {\mathbf 0}_{(N-k)\times (N-k) }.
\ee
one  obtains precisely the form dictated by conformal invariance \eqref{1-pointCFT} with 
\be\label{hweak}
{\mathbf h}^{(weak)}=-\frac{2}{g_{YM}^2} \frac{1}{12}\frac{(k^2-1)k}{8}=-\frac{1}{48 g_{YM}^2} (k^2-1)k.
\ee
Notice the difference of equations \eqref{normm}, \eqref{tmatr} and \eqref{comm} with respect to the analogous relations in the D3-D5 system.
\subsection{One-point function of the CPOs}\label{CPO-weak}
We now turn to the weak coupling calculation of CPOs. The symmetry of our D5-brane solution dictates that only operators respecting the $SO(3)\times SO(3)$ R-symmetry can have non-zero one-point functions. Thus, our case resembles that of the D3-D5 system. The relative spherical harmonic can be found in \eqref{sph}.
The most generic chiral primary operator of ${\cal N}=4$ SYM can be written as
\begin{equation}\label{chiralprimary}
{\cal O}_{\Delta I}(x)\equiv
\frac{(8\pi^2)^{\frac{\Delta}{2}}}
{\lambda^{\frac{\Delta}{2}}\sqrt{\Delta} }C_I^{i_1 i_2\ldots i_{\Delta}}
\,{\rm Tr}\left(\varphi_{i_1}(x)\varphi_{i_2}(x)\ldots \varphi_{i_{\Delta}}(x)\right),
\end{equation}
where the $\varphi_i$ denotes any of the six real scalar fields of the theory.
$\Delta$ is the conformal dimension of the operator which is protected by supersymmetry and as a result it does not depend on the 't Hooft coupling. 
The tensors $C_{I}^{i_1i_2\ldots i_{\Delta}}$ satisfy the condition
\begin{equation}\label{tracelesssymmetric}
\sum_{i_1\ldots i_\Delta=1}^6C_{I_1}^{i_1i_2\ldots i_{\Delta}}C_{I_2}^{i_1i_2\ldots i_{\Delta}}=\delta_{I_1I_2}~~.
\end{equation} 
and are symmetric
and traceless in the indices $i_1\ldots i_\Delta$.
Furthermore, the indices $I_a$ are there to label different such tensors.
The normalization (\ref{tracelesssymmetric}) implies that 
the planar limit of two point functions of the operators in (\ref{chiralprimary}) 
are normalized as follows
\begin{equation}\label{chiralprimarynormalization}
\langle
{\cal O}_{\Delta_1 I_1}(x){\cal O}_{\Delta_2I_2}(y)
\rangle=\frac{\delta_{I_1I_2}\delta_{\Delta_1\Delta_2}}{|x-y|^{2\Delta_1}}.
\end{equation}
Using the tensors appearing in (\ref{chiralprimary}), the spherical harmonics of $S^5$ can be written as
\begin{equation}
Y_{\Delta I}=C_I^{i_1 i_2\ldots i_{\Delta}} \hat x_{i_1}\hat x_{i_2}\ldots \hat x_{i_\Delta},
\label{tensor}
\end{equation}
where the $\hat x_i$'s are the components of a unit vector satisfying
\begin{equation}
\hat x_1^2+\hat x_2^2+\ldots +\hat x_{6}^2=1.
\end{equation} 
To make contact with \eqref{sph} we notice that $\hat x_1^2+\hat x_2^2+\hat x_3^2=\sin^2{\tilde\psi}$ while $\hat x_4^2+\hat x_5^2+\hat x_6^2=\cos^2{\tilde\psi}$. 

 The classical solution \eqref{sol} should now be substituted into the trace of equation (\ref{chiralprimary}). As mentioned above, the only components
which will contribute will be those which are invariant under the $SO(3)\times SO(3)$ subgroup
of the $SO(6)$ R-symmetry group. Consequently, only the chiral primary operators respecting this symmetry
will have non-vanishing one-point functions. As discussed in \cite{Kristjansen:2012tn}, for each even $\Delta=2 l$ there is a  unique $SO(3)\times SO(3)$ symmetric chiral
primary operator  which we denote by ${\mathcal O}_\Delta(x')$. The corresponding spherical harmonic is dnoted by $Y_\Delta$. Its normalisation will be chosen to be that of equation (\ref{chiralprimarynormalization}). 

In order to make explicit the presence of the spherical harmonic $Y_\Delta$ in the trace of (\ref{chiralprimary}) one should factor the normalisation of the six-dimensional vector $(\phi_1,\ldots,\phi_6)$ so that inside the trace one has the components of a unit
vector, as in \eqref{tensor}. From \eqref{normm} it is easy to see that this normalisation is $\frac{1}{2 \sqrt{2} r'}(k^2-1)^{1/2} $, so that we get an overall factor of the normalization to the power of the number of fields, i.e, the one-point function will be proportional to $\frac{(k^2-1)^{\Delta/2}}{(2 \sqrt{2} r')^\Delta}$. The final trace will give a factor of $k$.
Assembling everything together one obtains for the one-point function of the CPOs
\be\label{CPOweak}
\langle {\mathcal O}_\Delta(x')\rangle^{(weak)}=\frac{(8\pi^2)^{\frac{\Delta}{2}}}{\lambda^{\frac{\Delta}{2}}\sqrt{\Delta} }\frac{(k^2-1)^{\Delta/2}k}{(2 \sqrt{2} r')^\Delta}Y_\Delta(0)
\ee
This is our final result for the one-point function of CPOs in the weak coupling regime.

Before closing this section, let us make a brief comment concerning the integrability of our construction at weak coupling. Looking at the structure of the classical solution \eqref{sol} and if one restricts to the 
$SO(6)$ subsector it is obvious that the tree-level one-point function of a scalar operator will be given by the same determinant formula that is valid for the case of the supersymmetric D3-D5 defect \cite{deLeeuwKristjansenZarembo15, Buhl-MortensenLeeuwKristjansenZarembo15, deLeeuwKristjansenLinardopoulos18a, deLeeuwGomborKristjansenLinardopoulosPozsgay19}. Furthermore, the matrix product state (MPS) will also be the same as in the supersymmetric D3-D5 defect and will be annihilated by all odd charges $Q_{2n+1},\, n\in {\mathbb N}^*$. Thus, one is tempted to conjecture that the theory exhibits integrability at the weak coupling regime. However, it is not obvious to us that this similarity will continue to hold at the loop level.

\section{Agreement between the weak and strong coupling results}\label{agreement}

In this section, we will compare the weak and strong coupling results for both the one-point function of the energy-momentum tensor and the one-point function of CPOs calculated in the previous sections. We will see that in the appropriate limit, in which $\frac{k}{\sqrt{\lambda}}=\frac{\kappa}{\pi}\gg 1$ agreement is observed between the leading term in the expressions for the weak and strong coupling. This consists a highly non-trivial test of the correspondence. A similar agreement was also found in \cite{Nagasaki:2012re} and \cite{Kristjansen:2012tn} for the CPOs in the codimension-1 supersymmetric D3-D5 and non-supersymmetric D3-D7 systems, respectively. The situation here is reminiscent of the BMN limit, one engineers a quantity, here $\frac{\lambda}{k^2}$ in the BMN limit $\frac{\lambda}{J^2}$, which can be taken to be small at both weak and strong coupling. Then one compares the observables order by order in this small quantity. For the supersymmetric codimension-2 D3-D3 defect the analogous computations can be found in \cite{Gomis:2007fi,Drukker:2008wr}.
 
We start with the case of the one-point function of the energy-momentum tensor. The weak coupling result \eqref{hweak} agrees with that obtained at strong coupling \eqref{hT}.
To see this, one should keep in mind that the two results should be compared in the limit where 
\be\label{limit}
\frac{k}{\sqrt{\lambda}}=\frac{\kappa}{\pi}\gg 1\, . 
\ee
By looking at \eqref{A}, one sees that this happens in the limit where $\s\rightarrow 0$. In this limit, $k=\frac{\sqrt{2 \lambda}}{\pi \sigma}$ and the constant ${\mathbf h}^{(weak)}$ of \eqref{hweak} becomes
\be\label{hweak-1}
{\mathbf h}^{(weak)}=-\frac{\sqrt{2}}{24 g_{YM}^2}\frac{\lambda^{3/2}}{\pi^3 \s^3}\, .
\ee
On the other hand, in the same limit \eqref{hT} also becomes
\be\label{hT-1}
{\mathbf h}^{(strong)}=-\frac{\sqrt{2}}{24 g_{YM}^2}\frac{\lambda^{3/2}}{\pi^3 \s^3}\, 
\ee
which agrees with the value of ${\mathbf h}^{(weak)}$ at weak coupling.
The same agreement holds, obviously, for the B-type Weyl anomaly coefficient $d_2$ since the latter is related to the coefficient ${\mathbf h}$ through \eqref{hd2rel}.

We now turn to the case of the CPOs' one-point functions. In the same limit \eqref{limit}, the weak and strong coupling results also agree. Here we present the cases for $\D=4,6,8$ but we have checked that the agreement persists at least up to $\D=40$. It would be nice to prove this agreement for a general value of $\D=2 l,\, l\in {\mathbb N}^*$. The main obstacle here is the evaluation of the integrals \eqref{O-general} for generic $\D$.
Expanding \eqref{one-point-function-Delta-4-6-8} and \eqref{CPOweak} near $\sigma =0$, we arrive at the following expressions
\begin{eqnarray} \label{one-point-function-Delta-4-6-8-expansion}
  \langle {\cal O}_4(r')\rangle^{(weak)} = \langle {\cal O}_4(r')\rangle^{(strong)}& = & \sqrt{\frac{3}{10}} \, \frac{\sqrt{\lambda }}{\pi \, \sigma^5} \, \frac{1}{r'^4} \, +\cdots \nonumber \\[5pt]
   \langle {\cal O}_6(r')\rangle^{(weak)} = \langle {\cal O}_6(r')\rangle^{(strong)} & = &
   -\frac{2}{\sqrt{21}} \, \frac{\sqrt{\lambda }}{\pi \, \sigma^7} \, \frac{1}{r'^6} \, + \cdots
   \\[5pt]
    \langle {\cal O}_8(r')\rangle^{(weak)} = \langle {\cal O}_8(r')\rangle^{(strong)} & = & 
   \frac{\sqrt{5}}{6}\frac{\sqrt{\lambda }}{\pi \, \sigma^9}\, \frac{1}{r'^8} \, +\cdots 
    \nonumber\,\,\, ,
\end{eqnarray}
where the dots denote subleading terms in the large $\frac{k}{\sqrt{\lambda}}\sim \frac{1}{\s}$ expansion and which generically do not agree between weak and strong coupling.

In fact, the leading term of the one-point functions for CPOs of generic dimension $\D$ can be written in the limit \eqref{limit}. It reads
\begin{eqnarray}\label{genericD}
\langle {\cal O}_\Delta(r')\rangle^{(weak)} = 
\langle {\cal O}_\Delta(r')\rangle^{(strong)}& =  
(-1)^{\frac{\Delta}{2}} \, \sqrt{\frac{\Delta +2}{\Delta \left(\Delta +1\right)}} \, \frac{\sqrt{\lambda }}{\pi \, \sigma^{\Delta +1}} \, 
\frac{1}{r'^\Delta}  =  \nonumber \\
& =(-1)^{\frac{\Delta}{2}} \, \sqrt{\frac{\Delta +2}{\Delta \left(\Delta +1\right)}} \, \frac{\pi^{\D} \, k^{\D+1} }{2^{(\D+1)/2} \lambda^{\D/2}}  \, 
\frac{1}{r'^\Delta}.  
\end{eqnarray}
The expression above has been derived by expanding the weak coupling formula \eqref{CPOweak} and by using the relation $k=\frac{\sqrt{2 \lambda}}{\pi \sigma}$. We have checked that \eqref{genericD} agrees with the corresponding term in the strong coupling formula \eqref{O-general} up to $\D=40$.

\section{Conclusions}\label{concl}
In this work, we have initiated the study of a new holographic duality between a certain non-supersymmetric defect conformal field theory (dCFT) and its gravity dual. 

On the field theory side, the defect is a flat two-dimensional plane, that is $\mathbb{R}^{(1,1)}$ or $\mathbb{R}^{2}$ depending on whether the geometry of the embedding space is Lorentzian or Euclidean. The defect is embedded in the maximally supersymmetric field theory in 4-dimensions, $\mathcal N=4$ SYM. As usual, this codimension-2 defect induces singularities in the scalar fields of the ambient CFT. 
The 2-dimensional conformal interface is described by a classical solution whose precise form we determined.

On the gravity side, the defect is realised by a novel solution of a D5 probe brane embedded in the $AdS_5\times S^5$ geometry. The solution depends on a single parameter $\s$. For the solution to exist, $\s$ should be smaller than or equal to the value $2 \sqrt{2}$. This implies that the angle between the brane and the boundary of the space $\partial AdS_5$ should not exceed a certain critical value.
The D5 brane wraps an $S^2\subset S^5$ and carries $k$ units of flux through the $S^2$. The symmetry of the induced on the brane metric is $AdS_3\times S^1\times S^2$. The brane ends on an $\mathbb{R}^{(1,1)}$ subspace of the 4-dimensional boundary, resulting in a codimension-2 defect. 

We first examined the stability of our brane configuration. By expanding the brane action to quadratic order in the fluctuations of the transverse to the brane coordinates we found that the corresponding masses respect the B-F bound, as long as the parameter $\s\le 1$.
Thus, we see that the requirement of stability introduces a stricter bound than the equations of motion.
Subsequently, we holographically calculated the one-point function of the energy-momentum tensor, as well as that of the chiral primary operators (CPOs) by invoking the prescription that is vividly articulated in \cite{Georgiou:2023yak}. The corresponding calculations at the weak coupling regime were performed using the classical solution that is dual to our $D5$ probe brane \eqref{sol}. In an appropriate limit, we found compelling agreement between the weak and strong coupling results. This agreement provides strong evidence in favour of the proposed duality.
Finally, we also extracted one of the B-type Weyl anomaly coefficients, namely $d_2$, by exploiting its relation to the one-point function of the energy-momentum tensor. The weak/strong coupling agreement observed for the energy-momentum tensor obviously holds also for the B-type anomaly coefficient $d_2$.

Our work opens several directions for future research. One possible direction is to calculate the remaining anomaly coefficients associated with our defect CFT, namely $b$ and $d_1$ of \eqref{eq:defect-A}. Another interesting question is that regarding the integrability of our construction, at both the weak and strong coupling regimes. Finally, one may compute the one-loop corrections to the one-point functions of the energy-momentum tensor and CPOs at both weak and strong coupling in order to check if the agreement found at the leading order persists.

\subsection*{Acknowledgements}

G.L.\ was supported by the Research Start-up Fund of the Shanghai Institute for Mathematics and Interdisciplinary Sciences (SIMIS) and the National Development Research and Innovation Office (NKFIH) research grant K134946. The work of G.L.\ was supported in part by the National Research Foundation of Korea (NRF) grant funded by the Korea government (MSIT) (No.\ 2023R1A2C1006975), as well as by an appointment to the JRG Program at the APCTP through the Science and Technology Promotion Fund and Lottery Fund of the Korean Government. 
G.L. thanks the participants of the joint program [APCTP-2025-J01] 
held at APCTP, Pohang, South Korea for fruitful discussions.
This paper has been financed by the funding programme ``MEDICUS", of the University 
of Patras (D.Z. with grant number: 83800).

\appendix\section[Supersymmetry]{Supersymmetry \label{Appendix:Supersymmetry}}
In this Appendix,  we study the Poincare and conformal supersymmetries  preserved by  the solution \eqref{sol}. The aforementioned symmetries are generated by two ten dimensional Majorana-Weyl spinors  $\epsilon_1$ and  $\epsilon_2$ of opposite chirality.
One can determine the  supersymmetries  that are left unbroken by \eqref{sol}  
by studying the supersymmetry variation of the gaugino. The number of independent spinors for which the supersymmetry variation of the gaugino is zero is the number of the preserved (super)-conformal symmetries.

The metric of the CFT is given by:
\noindent
\be\label{minkowskico}
ds^2=-(dx^0)^2+(dx^1)^2+(dx^2)^2+(dx^3)^2=-(dx^0)^2+(dx^1)^2+dr^2+r^2 d\psi^2  .
\ee
A Poincare supersymmetry transformation is  given by 
\begin{eqnarray}\label{susyvar}
\delta \psi= \Big({1\over 2}F_{\mu\nu}\Gamma ^{\mu\nu}+D_{\mu}\varphi_i\Gamma^{\mu\, i+3}-{i\over 2}[\varphi_{i},\varphi_{j}]\Gamma^{i+3\,\,j+3}\Big) \epsilon_{1}
\end{eqnarray}
Here $\mu$ runs from 0 to 3 while $i$ runs from 1 to 6. 
$\Gamma ^{\mu}$ and $\Gamma ^{i+3}$ denote the 10-dimensional gamma matrices satisfying 
$\{\Gamma_M,\Gamma_N\}=-2 \,\eta_{MN}{\mathbf 1}_{10}$, where $M,N=0,1,\dots,9$ and $\eta_{MN}$ the mostly plus flat metric in 10 dimensions. By inserting \eqref{sol} in \eqref{susyvar} we obtain
\begin{eqnarray}
\delta \psi=-\frac{x'_2}{r'^3}\,\Gamma^{2, i+6}\epsilon_{1}\otimes {\cal T}_i -\frac{x'_3}{r'^3}\,\Gamma^{3, i+6}\epsilon_{1}\otimes{\cal T}_i  
+\frac{1}{2\sqrt{2} \,r'^2}\,\epsilon_{ijl}\, \Gamma^{i+6,j+6}\epsilon_{1}\otimes{\cal T}_l,
\end{eqnarray}
where $i=1,2,3$ and ${\cal T}_i=\left[\begin{array}{cc} \left(t_i\right)_{k\times k} & 0_{k\times \left(N - k\right)} \\ 0_{\left(N - k\right)\times k} & 0_{\left(N - k\right)\times \left(N - k\right)} \end{array}\right]_{N\times N}$. By focusing, for example,  on the term proportional to $\frac{x'^2}{r'^3}$ we get
\be
\delta \psi=0\quad  \Rightarrow \quad \Gamma^{2,i+6}\epsilon_{1}=0 \quad  \Rightarrow \quad
\epsilon_{1}=0
\ee
since the matrix $\Gamma^{2,i+6}$ is invertible. We conclude that none of the supersymmetries of the $AdS_5\times S^5$ background is preserved in the presence of the defect.

Similarly, the superconformal supersymmetry transformations are given by
\begin{eqnarray}\label{supercon}
\delta \psi&=\Big( ({1\over 2}F_{\mu\nu}\Gamma ^{\mu\nu}+D_{\mu}\varphi_i\Gamma^{\mu\, i+3}-{i\over 2}[\varphi_{i},\varphi_{j}]\Gamma^{i+3\,\,j+3}) x^{\sigma}\Gamma _{\sigma}-2\varphi_{i}\Gamma^{i+3}\Big) \epsilon_{2}
\end{eqnarray}
and by focusing on the expression originating from the last term in \eqref{supercon}
one gets
\be
\delta \psi=0\quad  \Rightarrow \quad \Gamma^{i+6}\epsilon_{2}=0 \quad  \Rightarrow \quad
\epsilon_{2}=0, \, \quad \textrm{with } \quad i=1,2,3.
\ee
As a result nor any of the superconformal symmetries is preserved. 
These results are in agreement to the strong coupling analysis. There the presence of the D5 probe brane also breaks all sypersymmetries.


\section{Details of the computations}
\label{Appendix:details_computations}

In this Appendix we gather useful details of the computations that are presented in the main text. 

The determinant of the induced metric on the D5-brane 
\be\label{det}
\sqrt{h}=\frac{ \sqrt{\left(\kappa ^2+1\right) \left(\sigma ^2+1\right)}}{r^3 \sigma ^4}\sin{\beta}
\ee
and the quantity
\begin{eqnarray} \label{D-brane-insertion}
&&h^{ab}\partial_a Y^\mu \partial_b Y^\nu \frac{\partial x'^{\kappa}}{\partial x^\mu} \frac{\partial x'^{\lambda}}{\partial x^\nu}=\nonumber\\
&&\begin{pmatrix}
r^2 \, \sigma^2 & 0 & 0 & 0 & 0 \\[5pt]
0 & r^2 \, \sigma^2  & 0 & 0 & 0  \\[5pt]
0 & 0 &\frac{r^2 \sigma ^2 \left((1+\sigma ^2) \sin ^2{\psi} +\cos ^2{\psi} \right)}{1+\sigma ^2} & -\frac{r^2 \sigma ^4 \sin{\psi}  \cos{\psi} }{1+\sigma ^2} & \frac{r^2 \sigma ^3 \cos{\psi} }{1+\sigma ^2} \\[5pt]
0 & 0 &  -\frac{r^2 \sigma ^4 \sin{\psi}  \cos{\psi} }{1+\sigma ^2}  & r^2 \sigma ^2 (\frac{\sin^2\psi }{1+\sigma ^2}+\cos ^2\psi ) & \frac{r^2 \sigma ^3 \sin{\psi} }{1+\sigma ^2}  \\[5pt]
0 & 0 &  \frac{r^2 \sigma ^3 \cos{\psi} }{1+\sigma ^2} & \frac{r^2 \sigma ^3 \sin{\psi} }{1+\sigma ^2} & \frac{r^2 \sigma ^4}{1+\sigma ^2}
\end{pmatrix}
\end{eqnarray}
are needed for the calculation of the one-point function of the stress tensor in section \ref{Tmunu}.

For the calculation of the one-point function of chiral primary operators, which is presented in section \ref{CPOs}, we need several ingredients to be substituted in \eqref{Fluct_DBI+WZ}. The induced metric on the D5-brane is 
\begin{equation}
H = \begin{pmatrix}
\frac{1}{r^2 \, \sigma^2} & 0 & 0 & 0 & 0 & 0 \\[5pt]
0 & \frac{1}{r^2 \, \sigma^2}  & 0 & 0 & 0 & 0 \\[5pt]
0 & 0 & \frac{1+\sigma^2 }{r^2 \, \sigma^2}  & 0 & 0 & 0\\[5pt]
0 & 0 & 0 & \frac{1}{\sigma^2} & 0 & 0\\[5pt]
0 & 0 & 0 & 0 & 1 & \kappa \, \sin\beta \\[5pt]
0 & 0 & 0 & 0 & - \kappa \, \sin\beta & \sin^2 \beta
\end{pmatrix}	
\end{equation}
and the symmetric part of $H^{-1}$ is 
\begin{equation}
H^{-1}_{sym} = 
\begin{pmatrix}
    r^2 \, \sigma^2 & & & & & \\
   & r^2 \, \sigma^2  & & & & \\
   & & \frac{r^2 \, \sigma^2}{1+\sigma^2} & & & \\
   & & & \sigma^2 & &\\
   & & & &  \frac{1}{1+\kappa^2} & \\
   & & & & & \frac{\sin^{-2}\beta}{1+\kappa^2}
\end{pmatrix} \, . 
\end{equation}
The components of the fluctuation of the metric that are needed for the computation of the DBI contribution are
\begin{eqnarray} \label{metric_fluctuations}
\frac{h_{00}}{\Delta \, s} & = & - \frac{2}{z^2} +16 \frac{\left(x_0-X_0\right)^2}{K^2} \, , \quad 
\frac{h_{11}}{\Delta \, s} =  - \frac{2}{z^2} +16 \frac{\left(x_1-X_1\right)^2}{K^2} 
\nonumber \\ 
\frac{h_{22}}{\Delta \, s} & = & - \frac{2}{z^2} +16 \frac{\left(r - r' \, \cos (\psi - \psi')\right)^2}{K^2} \, , \quad 
\frac{h_{33}}{\Delta \, s} =  - \frac{2}{z^2} +16 \frac{\left(r \, r'  \, \sin (\psi - \psi') \right)^2}{K^2} 
\nonumber \\ 
\frac{h_{24}}{\Delta \, s} & = & 16 \, z \, \frac{r - r' \, \cos (\psi - \psi')}{K^2} - 
 8 \, \frac{r - r' \, \cos (\psi - \psi')}{z\, K} 
 \nonumber \\ 
\frac{h_{44}}{\Delta \, s} & = &  \frac{2}{z^2} - \frac{16}{K} +\frac{16 \, z^2}{K^2} \, , \quad
\frac{h_{88}}{\Delta \, s} =  2 \, , \quad \frac{h_{99}}{\Delta \, s} =  2 \, \sin^2 \beta
\end{eqnarray}
where
\begin{equation}
K = z^2+\left(x_0-X_0\right)^2 +\left(x_1-X_1\right)^2 + r^2  + r'^2 - 2 \, r \, r' \, \cos (\psi - \psi')
\end{equation}
and
\begin{equation} \label{s_def}
s = c_{\Delta} \, \frac{z^{\Delta}}{K^{\Delta}} \, Y_{\Delta}(\tilde \psi)
\quad {\rm with} \quad c_{\Delta} = \frac{\Delta +1 }{2^{2 -\frac{\Delta}{2}}\, N \, \sqrt{\Delta}} \, . 
\end{equation}
In \eqref{metric_fluctuations}, $(X_0,X_1,r',\psi',0)$ is the position of the boundary where the CPO is located while  $(x_0,x_1,r,\psi,z)$ is an arbitrary point on the D5 brane. Furthermore, $z=\s\, r$ since the equations above are evaluated on the brane.


\section{Graviton propagator}
\label{Appendix:graviton-fluct}
Here we comment on a subtlety concerning the bulk-to-boundary graviton propagator.
From \cite{Arutyunov:1999nw} the fluctuation of the bulk metric due to a disturbance on the boundary of the $AdS_5$ space is given by
\begin{IEEEeqnarray}{l}
\delta g_{mn}\left(y,w\right) = \int d^4x' \,\frac{d+1}{d-1} \cdot \frac{1}{w^2} \times \K_d(y,w;x) \, 
\I_{m k}\left(\tilde r\right) \, \I_{n l}\left(\tilde r\right)  \, \I_{kli'j'}\, \delta g_{i'j'}(x')\, ,
\label{Gravitonfluctuations}
\end{IEEEeqnarray}
where $d=4$ and the various quantities in the formula above are defined in section \ref{Tmunu}. Let us also remind the reader that $m,n=0,1,\dots, 4$ while $i,j=0,1,\dots, 3$.
Equation \eqref{Gravitonfluctuations} should now be plugged in $\big\langle\delta g_{ij}\left(x\right) \cdot\delta g_{mn}\left(y,w\right)\big\rangle$ in order to obtain the graviton propagator. To this end one should use the following overlap for the boundary fluctuations of the metric
\be\label{overlap}
\big\langle\delta g_{ij}\left(x\right) \cdot\delta g_{i'j'}\left(x'\right)\big\rangle=(\d_{ii'}\d_{jj'}+\d_{ij'}\d_{ji'})\, \d^{(4)}(x-x')\, .
\ee
Now the fact that the tensor $ \I_{kli'j'}$ is symmetric in its last two indices gives the factor of 2 in front of the right hand side of \eqref{GravitonPropagator}. 




\bibliographystyle{utphys}

\bibliography{refs}

\end{document}